\documentclass[twocolumn]{aastex631}

\usepackage[super]{nth}
\usepackage{hyperref}

\begin{document}

\title{
Star formation and AGN activity 500 Myr after the Big Bang: Insights from JWST
}

\author[0000-0002-9816-1931]{Jordan C. J. D'Silva}
\affiliation{International Centre for Radio Astronomy Research (ICRAR) and the
International Space Centre (ISC), The University of Western Australia, M468,
35 Stirling Highway, Crawley, WA 6009, Australia}
\affiliation{ARC Centre of Excellence for All Sky Astrophysics in 3 Dimensions (ASTRO 3D), Australia}

\author[0000-0001-9491-7327]{Simon P. Driver} 
\affiliation{International Centre for Radio Astronomy Research (ICRAR) and the
International Space Centre (ISC), The University of Western Australia, M468,
35 Stirling Highway, Crawley, WA 6009, Australia}

\author[0000-0003-3021-8564]{Claudia D. P. Lagos}
\affiliation{International Centre for Radio Astronomy Research (ICRAR) and the
International Space Centre (ISC), The University of Western Australia, M468,
35 Stirling Highway, Crawley, WA 6009, Australia}
\affiliation{ARC Centre of Excellence for All Sky Astrophysics in 3 Dimensions (ASTRO 3D), Australia}

\author[0000-0003-0429-3579]{Aaron S. G. Robotham} 
\affiliation{International Centre for Radio Astronomy Research (ICRAR) and the
International Space Centre (ISC), The University of Western Australia, M468,
35 Stirling Highway, Crawley, WA 6009, Australia}

\author[0000-0002-7265-7920]{Jake Summers} 
\affiliation{School of Earth and Space Exploration, Arizona State University,
Tempe, AZ 85287-1404, USA}

\author[0000-0001-8156-6281]{Rogier A. Windhorst}
\affiliation{School of Earth and Space Exploration, Arizona State University,
Tempe, AZ 85287-1404, USA}



\begin{abstract}

We consider the effect of including an active galactic nuclei (AGN) component when fitting spectral energy distributions of 109 spectroscopically confirmed $z\approx 3.5-12.5$ galaxies with \textit{JWST}. Remarkably, we find that the resulting cosmic star formation history is $\approx 0.4$~dex lower at $z\gtrsim 9.5$ when an AGN component is included in the fitting. This alleviates previously reported excess star formation at $z\gtrsim 9.5$ compared to models based on typical baryon conversion efficiencies inside dark matter halos. We find that the individual stellar masses and star formation rates can be as much as $\approx 4$~dex lower when fitting with an AGN component. These results highlight the importance of considering both stellar mass assembly and supermassive black hole growth when interpreting the light distributions of among the first galaxies to ever exist. 

\end{abstract}




\section{Introduction} \label{sec:intro}
Star formation and the growth of supermassive black holes (SMBHs) are key ingredients in galaxy formation as evidenced by their major contributions to the light distributions of galaxies \citep{kauffmann_unified_2000, schaye_physics_2010}. Star formation manifests at many different wavelengths, from the ultraviolet all the way to the radio \citep{conroy_modeling_2013, davies_gamah-atlas_2016}. While SMBHs are unobservable, their contributions to spectral energy distributions (SEDs) originate from the accretion disk that surrounds the black hole, powering so-called active galactic nuclei (AGN) that radiate at similar wavelength regimes to the star formation \citep{richards_sloan_2006,shen_bolometric_2020}. Analyses, and interpretations, of star formation in galaxies may therefore be contaminated by AGN because these processes manifest at similar wavelengths and with potentially similar contributions to the total SED of galaxies \citep{cardoso_impact_2017}. 

Thus, galaxies with a significant AGN component are often excluded from studies of star formation. Indeed, this was the approach taken by \citet{driver_gamag10-cosmos3d-hst_2018} in calculating the cosmic star formation history \citep[CSFH;][]{lilly_canada-france_1996,madau_cosmic_2014} where they used a combination of the mid-infrared and radio luminosities to effectively filter out strong AGN from their sample, finding that only a $\approx 0.1$~dex uncertainty is induced on the CSFH out to $z\approx 5$ as a result of the AGN filtering. While it is encouraging that this selection does not precipitate a greater uncertainty, this approach fails to acknowledge the coevolution of star formation and the growth of SMBHs. 

In fact, AGN are thought to inject energy into the host galaxy through the process of accretion-driven feedback. The net effect is to enhance the velocity dispersion of the surrounding gas and/or drive gas outflows \citep{magorrian_demography_1998, ferrarese_fundamental_2000} that can affect star formation \citep{davies_galaxy_2019,katsianis_evolving_2019,matthee_origin_2019, davies_deep_2022}. In light of these results, if we are to understand how galaxies form and evolve it is crucial to consider the union of these two fundamental processes. 

This is especially true at the $z\approx 3.5-12.5$ frontier currently being probed by the \textit{James Webb Space Telescope} (\textit{JWST}), among whose key objectives is to shed light on the formation of the first stars and black holes in the first galaxies \citep{gardner_james_2006}. Indeed, \textit{JWST} has already provided valuable insights into the first galaxies and AGN \citep[e.g.,][]{adams_discovery_2023,yan_jwsts_2023,curtis-lake_spectroscopic_2023,haro_spectroscopic_2023,naidu_two_2022,bouwens_uv_2023, bunker_jades_2023,larson_ceers_2023,kocevski_hidden_2023, juodzbalis_epochs_2023,ubler_ga-nifs_2023}. 

At this stage, \textit{JWST} has arguably generated more questions than it has answered about the hitherto unexplored early Universe. Recent results suggest that star formation was more efficient at $z\gtrsim10$ than predicted by models based on typical baryon conversion efficiencies inside dark matter halos \citep[e.g.,][]{harikane_goldrush_2022,harikane_comprehensive_2023}. \textit{JWST} has also yielded candidates of quenched galaxies at $z\approx 3$ whose star formation histories suggest a burst of star formation at $z\approx 11$ that appears to be at odds with our current grasp of stellar mass assembly \citep[e.g.,][]{carnall_surprising_2023,glazebrook_extraordinarily_2023}. 

In this work, we investigate these emergent puzzles by exploring the union of star formation and AGN activity at $z\gtrsim 3$. We use the SED fitting code \textsc{ProSpect} \citep{robotham_prospect_2020} to determine the physical properties of $z\approx 3.5-12.5$ galaxies. Critically, \textsc{ProSpect} is able to account for the contribution from both stars and the AGN when fitting the galaxy SED, and thus an invaluable tool to understand the interface of star formation and black hole growth. We build upon the analysis of \citet{dsilva_gamadevils_2023} and investigate the effect of AGN on the CSFH. 

In Section~\ref{sec:data} we describe our \textit{JWST} data and high-redshift galaxy sample. In Section~\ref{sec:methods} we describe our methods of SED fitting and calculating the CSFH. In Section~\ref{sec:results} we present our main results. Our conclusions are summarised in Section~\ref{sec:conclusion}. We use concordance $\Lambda$CDM cosmology throughout, with $H_{0} = 70 \, \mathrm{km \, s^{-1} \ Mpc^{-1}}$, $\mathrm{\Omega_{M} = 0.3}$ and $\mathrm{\Omega_{\Lambda}=0.7}$. All of our results are derived using a \citet{chabrier_galactic_2003} initial mass function. 

\section{Data} \label{sec:data}

\subsection{Spectroscopic sample} \label{subsec:sample}
Our high-redshift sample is drawn from two \textit{JWST} early release science (ERS) programs: (i). the Cosmic Evolution Early Release Science survey \citep[CEERS;][ERS program 1345]{finkelstein_ceers_2023} and (ii). the GLASS-\textit{JWST} survey \citep[][ERS program 1324]{treu_glass-jwst_2022}. 

We used galaxies with confirmed spectroscopic redshifts with the NIRSpec instrument on \textit{JWST} that were presented in \citet{haro_spectroscopic_2023} and \citet{nakajima_jwst_2023}. Our sample includes the $z=12.17$ galaxy observed in the GLASS survey whose redshift was spectroscopically confirmed with the Atacama Large Millimetre Array \citep[ALMA;][]{bakx_deep_2023}. Our final sample comprises 109 galaxies with spectroscopic redshifts from $z\approx 3$ to $z\approx 12$, with 108 galaxies from the CEERS survey and the $z=12.17$ galaxy from GLASS. 

\subsection{Imaging and photometry} \label{subsec:photometry}
On the basis of our final spectroscopic sample, we made photometric catalogues suitable for SED fitting. Our photometric catalogues are derived from the combination of \textit{Hubble Space Telescope} (\textit{HST}) data from legacy surveys and \textit{Near Infrared Camera} \citep[NIRCam;][]{rieke_overview_2005,rieke_performance_2023} \textit{JWST} imaging. 

For the NIRCam data, all available \texttt{UNCAL} files from the two programs were downloaded from the Mikulski Archive for Space Telescopes (MAST) and processed through to the \texttt{CAL} stage using the \texttt{JWST Calibration Pipeline}\footnote{\url{https://github.com/spacetelescope/jwst}}. Our CEERS reductions used pipeline version 1.10.2 \citep{bushouse_jwst_1102_2023} and CRDS pmap=1084. Our GLASS reductions used pipeline version 1.11.2 \citep{bushouse_jwst_1112_2023} and pmap=1100. We do not expect significant differences between these data as a result of these slightly different calibration versions, especially since \textit{JWST's} NIRCam performance has significantly converged since its first light \citep[e.g.,][]{adams_discovery_2023}. As a test, we remade a subset of the CEERS data with the updated pipeline version and pmap, finding negligible differences in the imaging data for all NIRCam modules and filters between versions. The most significant update between pipeline versions is improved snowball and outlier detection. For reference, CRDS version 1084 was implemented on 2023 May 5 and CRDS version 1100 was implemented on 2023 July 14. 

We performed a custom pipeline step that corrects for $1/f$ noise and wisp artefacts and subtracts the sky background in the images. These custom steps were built on the source extraction tool \textsc{ProFound} \citep{robotham_profound_2018}. Further details on the $1/f$ removal can be found in \citet{windhorst_webbs_2022} and details on the wisp removal step can be found in \citet{robotham_dynamic_2023}. 

These artefact-free and sky-subtracted \texttt{CAL} files were then used to make mosaics via inverse-variance weighted stacking with the novel code \textsc{ProPane}\footnote{\url{https://github.com/asgr/ProPane}} \citep[for further details, please see][]{robotham_propane_2023}. Hot pixels were patched with the equivalent pixels from a median stack. We draw attention to this stacking step as it is an entirely novel stacking code, unlike many of the custom JWST pipelines (e.g., GRIZLI, PENCIL) that rely on DrizzlePac \footnote{\url{https://github.com/spacetelescope/drizzlepac}}.

\textit{HST} observations were folded in to improve our coverage of the SEDs. For CEERS we downloaded the \textit{HST} mosaics provided by the CEERS team, while for GLASS we downloaded Hubble Multivisit Mosaics that overlapped with our \textit{JWST} NIRCam observations. The CEERS mosaics are built from CANDELS observations in the Extended Groth Strip \citep{koekemoer_candels_2011}. The GLASS \textit{HST} mosaics are built from many legacy surveys \citep[e.g., Frontier Fields][]{lotz_frontier_2017} where all observations that overlap in cells projected on the sky are combined. These \textit{HST} observations are aligned to the \textit{JWST} observations using \textsc{ProPane} and the pixel scale is $0.06$~arcsec$/$pixel for all \textit{HST} and \textit{JWST} imaging.

Source detection was performed on an all-filter inverse-variance weighted stack of the NIRCam images, spanning a wavelength range of $\approx 1.0 \mu \mathrm{m} \to 4.5 \mu \mathrm{m}$. \textsc{ProFound} was used to create the initial source detection map and extract multiband measurements from each individual filter. \textsc{ProFound} has been used to make the multiband catalogues for the Galaxy and Mass Assembly (GAMA) Survey \citep{bellstedt_galaxy_2020-1} and the Deep Extragalactic VIsible Legacy Survey \citep[DEVILS;][]{davies_deep_2021}. 

Briefly, \textsc{ProFound} calculates a grid of sky statistics to model the sky in our detection image and flux peaks are then first identified if they are above some user-defined multiple of the sky RMS and number of connecting pixels (in our case we use $1.5\sigma_{\mathrm{skyRMS}}$ and seven connecting pixels). \textsc{ProFound} then searches for connecting pixels to grow the segmentation map for the sources. Overblending is mitigated through the use of `watershed' deblending in \textsc{ProFound} where flux peaks of neighbouring pixels need to be sufficiently distinct to be considered multiple sources (in our case blending only occurs if the saddle point in flux between two neighbouring sources is $<3\sigma_{\mathrm{skyRMS}}$ of the height of the highest flux peak). To capture all of the flux, \textsc{ProFound} then iteratively dilates the detected segments until the change in added flux with each dilation approaches zero. This differs from the commonly adopted methodology of \textsc{SourceExtractor} \citep[e.g.,][]{bertin_sextractor_1996} as the segments are not confined to a fixed shape (they conform to the natural morphology of the galaxy) and they capture most, if not all, of the flux, eschewing any need for aperture corrections.

Finally, to obtain realistic photometric errors, circular apertures of varying radii were placed in blank parts of each of our multiband images and the flux errors on similarly sized segments were scaled up by the value in the equivalent aperture. This step is important as spatial covariance introduced in the mosaicking stage is not captured by the Poisson error calculated inside of the galaxy segment. This sky sampling also allows us to account for cross terms in the error budget induced from detector and sky fluctuations. Additionally, we quadrature add a 10 per cent uncertainty on top of the scaled \textsc{ProFound} errors to further account for potential systematic errors. Ultimately, we aim to establish credible errors on our photometry to obtain robust constraints on the final physical parameters derived from SED fitting. All of our SEDs span from $\approx 0.6 \mu \mathrm{m} \to 4.5 \mu \mathrm{m}$. 

The full suite of processing, source detection and multiband photometry codes, tailored for \textit{JWST}, is referred to as \textsc{Jumprope} and is publicly available via \textsc{Github} \citep{jumprope}\footnote{\url{https://github.com/JordanDSilva/JUMPROPE}}.

\section{Methods} \label{sec:methods}

\subsection{\textsc{ProSpect} fitting} \label{subsec:prospect_fitting}
With our combined, multiwavelength catalogue we conducted SED fitting with \textsc{ProSpect} to determine physical quantities of the galaxies. A key advantage of \textsc{ProSpect} is its ability to distinguish the relative contributions to the SED from star formation and AGN.

Stellar emission is modelled by the \citet{bruzual_stellar_2003} templates with the \citet{chabrier_galactic_2003} initial mass function and the metallicity-dependent line energy tables from \citet{levesque_theoretical_2010} for nebular emission lines. Dust attenuation and reemission is modelled by the \citet{charlot_simple_2000} attenuation law and the \citet{dale_two-parameter_2014} far-infrared templates. Star formation histories (SFHs) are modelled parametrically as skewed normal distributions \citep[eqs. 1-5 in][]{dsilva_gamadevils_2023}. While parametric SFHs cannot capture random bursts of star formation, skewed normal distributions have been shown to follow the average shape of SFHs when applied to galaxy formation simulations \citep[e.g.,][]{robotham_prospect_2020}. A key advantage of \textsc{ProSpect} is its treatment of evolving metallicity, where metallicity is linearly mapped to the stellar mass growth and encapsulates chemical enrichment of the interstellar medium. The implementation of skewed normal parametric SFHs and physically motivated metallicity evolution in \textsc{ProSpect} is crucial for reproducing the correct CSFH when forensically constructed from $z\approx 0$ SFHs, suggesting that \textsc{ProSpect} accurately recovers the epochs at which stars formed \citep{bellstedt_galaxy_2020}. AGN are characterised by the models of \citet{fritz06agnmodel}.  Emission of the primary source is modelled as a combination of power laws. The dusty torus is modelled as a flared disk with a range of dust grain sizes. The opening angle and the angle of observation are also incorporated in the \citet{fritz06agnmodel} models. \textsc{ProSpect} has previously been shown to accurately recover the star-forming main sequence (SFMS), stellar mass function (SMF) and AGN bolometric luminosity function from $z\approx 0-5$ by e.g., \citet{thorne_deep_2021,thorne_deep_2022,dsilva_gamadevils_2023}. 

\textsc{ProSpect} was specifically designed to fit broadband SEDs of large samples; as such, only the \textit{HST} and \textit{JWST} photometry, and not the NIRSpec spectra, were fitted. The redshift was fixed to be the spectroscopic redshift of the sample to effectively eliminate potential redshift uncertainty on the fitted astrophysical properties. 

\begin{table}
    \centering
    \caption{Description of our two \textsc{ProSpect} runs. We refer to these descriptions throughout the text.}
    \label{tab:prospect_runs}

    \begin{tabular*}{\columnwidth}{m{3.5cm} m{4cm}}
    \textsc{ProSpect} Run Labels  & Description \\
    \hline
    \textsc{Stellar} & SFR and $\mathrm{M_{\star}}$ determined from the pure stellar component run of \textsc{ProSpect}, i.e., no AGN component was included to model the galaxy SEDs \citep[e.g.,][]{bellstedt_galaxy_2020,thorne_deep_2021}. \\
    \hline
    \textsc{Stellar+AGN} & SFR, $\mathrm{M_{\star}}$ and AGN bolometric luminosity of galaxies determined from the \textsc{ProSpect} run that included both AGN and stellar components \citep[e.g.,][]{thorne_deep_2022}. \\
    \hline
    \end{tabular*}
\end{table}
We used two flavours of \textsc{ProSpect} runs that are highlighted in Tab.~\ref{tab:prospect_runs}. Essentially, \textsc{Stellar+AGN} accounts for a stellar and AGN component in fitting the SED, while \textsc{Stellar} only accounts for a stellar component. The idea with using both of these SED fits is to bracket the extent to which an AGN may contribute to the SED. 

\begin{figure*}
\centering
    \includegraphics[width=\textwidth]{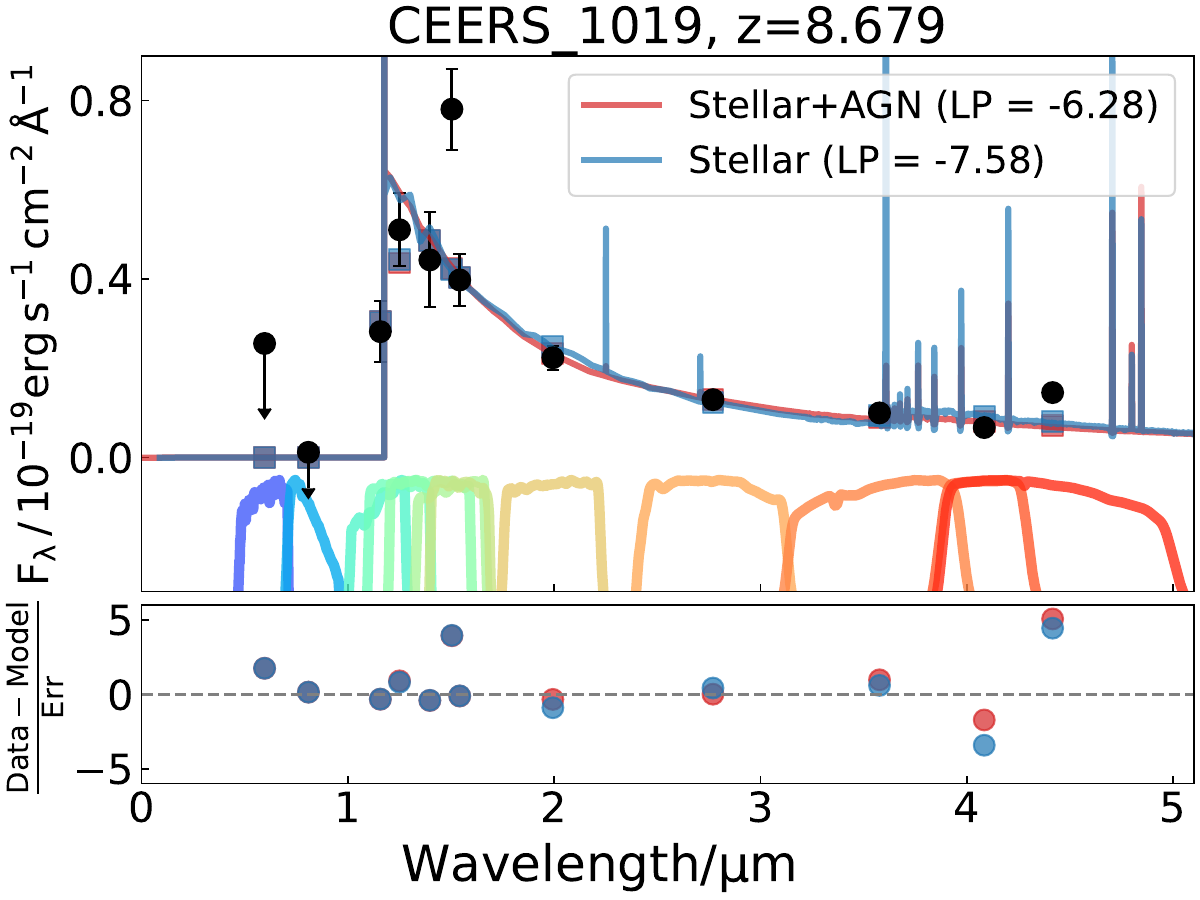}
    \caption{\textsc{ProSpect} fit to CEERS\_1019, a confirmed broad-line AGN at $z=8.679$ first presented in \citet{larson_ceers_2023}. The black points with error bars show the \textsc{ProFound} photometry. The error bars show the 16th-84th percentile ranges of the photometry. The \textsc{Stellar} fit is shown in blue, while the \textsc{Stellar+AGN} fit is shown in red. The log-probability (LP), being the maximum of the log-posterior distribution, is also included. The blue and red squares show the fit photometry for \textsc{Stellar} and \textsc{Stellar+AGN}. We show the relevant filter transmission curves from \textit{HST}+\textit{JWST} with rainbow colours. \citet{larson_ceers_2023} report $\mathrm{L_{AGN}\approx 10^{45.1 \pm 0.2} \, erg \, s^{-1}}$, while we find $\mathrm{L_{AGN}\approx 10^{45.02 \pm 2.35} \, erg \, s^{-1}}$. The bottom panel shows the error weighted difference between the observed and fitted photometry, with the same colour scheme used for the fitted SED curves above.}
    \label{fig:sed_fit}
\end{figure*}

Fig.~\ref{fig:sed_fit} shows both \textsc{Stellar+AGN} and \textsc{Stellar} fits to the confirmed broad-line AGN in CEERS first presented by \citet{larson_ceers_2023}. We see that a comparison of the maximum log-probability (LP) values shows that there is a factor of $10^{1.3}\approx 20$ times better fit probability when accounting for an AGN component. From the comparison of the observed photometry to the fit photometry in the bottom panel, there is overall good agreement highlighting that we are adequately fitting the SED of this galaxy. \citet{larson_ceers_2023} also provide estimates of the bolometric luminosity of the AGN, $\mathrm{L_{AGN}\approx 10^{45.1 \pm 0.2} \, erg \, s^{-1}}$, that is calculated from the broad component H$\alpha$ flux. Comparing this to our measurement from the multiband photometry, $\mathrm{L_{AGN}\approx 10^{45.02 \pm 2.35} \, erg \, s^{-1}}$, the agreement is remarkable. Although, we note that the uncertainty on the AGN bolometric luminosity from \textsc{Stellar+AGN} is much greater than that derived from the broad component of the H$\alpha$ line in \citet{larson_ceers_2023}. Ultimately, this consistency reassures us that \textsc{ProSpect} produces reliable SED fits when simultaneously accounting for AGN and stars. These \textsc{ProSpect} SED fits are then used to calculate the CSFH.

\subsection{Calculating CSFH} \label{subsect:calculating_csfh}
Usually, estimates of the CSFH rely on integration of the SFR distribution function or, as often is the case at high redshift, integration of the UV luminosity function as a proxy for SFR \citep[e.g.,][]{adams_epochs_2023,harikane_comprehensive_2023,bouwens_uv_2015}. This approach requires that the distribution function be well known with careful consideration of the sample selection and effective volumes \citep[e.g.,][]{weigel_stellar_2016}. Essentially, the idea is to account for all sources of star formation in the Universe and not underestimate the star formation by missing the faint sources, e.g., Malmquist bias. 

Unfortunately, the gain in using spectroscopically confirmed $z\gtrsim3.5$ galaxies, as opposed to photometric candidates, is somewhat offset by the loss in knowledge of the selection criteria used to observe those galaxies. It is difficult to directly compute the effective volumes of this sample and, as such, hard to compute the distribution function. This difficulty is highlighted in \citet{harikane_pure_2023} who use a similar spectroscopic sample to ours to estimate the UV luminosity function and CSFH. In this work, we instead pursued an alternative approach to calculate the SFR density distribution and CSFH. The logic is 

\begin{equation}
    \mathrm{ \rho_{SFR} } = \int ^ {\infty}_{-\infty}f( \mathrm{M_{\star}})\phi(\mathrm{M_{\star}})\mathrm{dM_{\star}},
    \label{eq:csfh}
\end{equation}

where $\phi(\mathrm{M_{\star}})$ is the SMF and $f(\mathrm{M_{\star}})$ is a function that describes the correlation between SFR and stellar mass. The SMF in Eq.~\ref{eq:csfh} is used only to obtain the effective volumes of the galaxies. Thus, in this work, $f(\mathrm{M_{\star}})$ is the independent variable observed between \textsc{Stellar+AGN} and \textsc{Stellar} \textsc{ProSpect} runs, and the effect on the CSFH when AGN is taken into account can be explored.

\subsubsection{Fitting the SFMS} \label{subsubsect:fitting_sfms}
Therefore, this alternative approach to derive the CSFH hinges on the existence of the SFMS, the tight sequence of SFR and stellar mass on which the bulk of galaxies tend to lie \citep{noeske_star_2007,salim_uv_2007,speagle_highly_2014}. 

\begin{figure*}
    \centering
    \includegraphics[width = 1\textwidth]{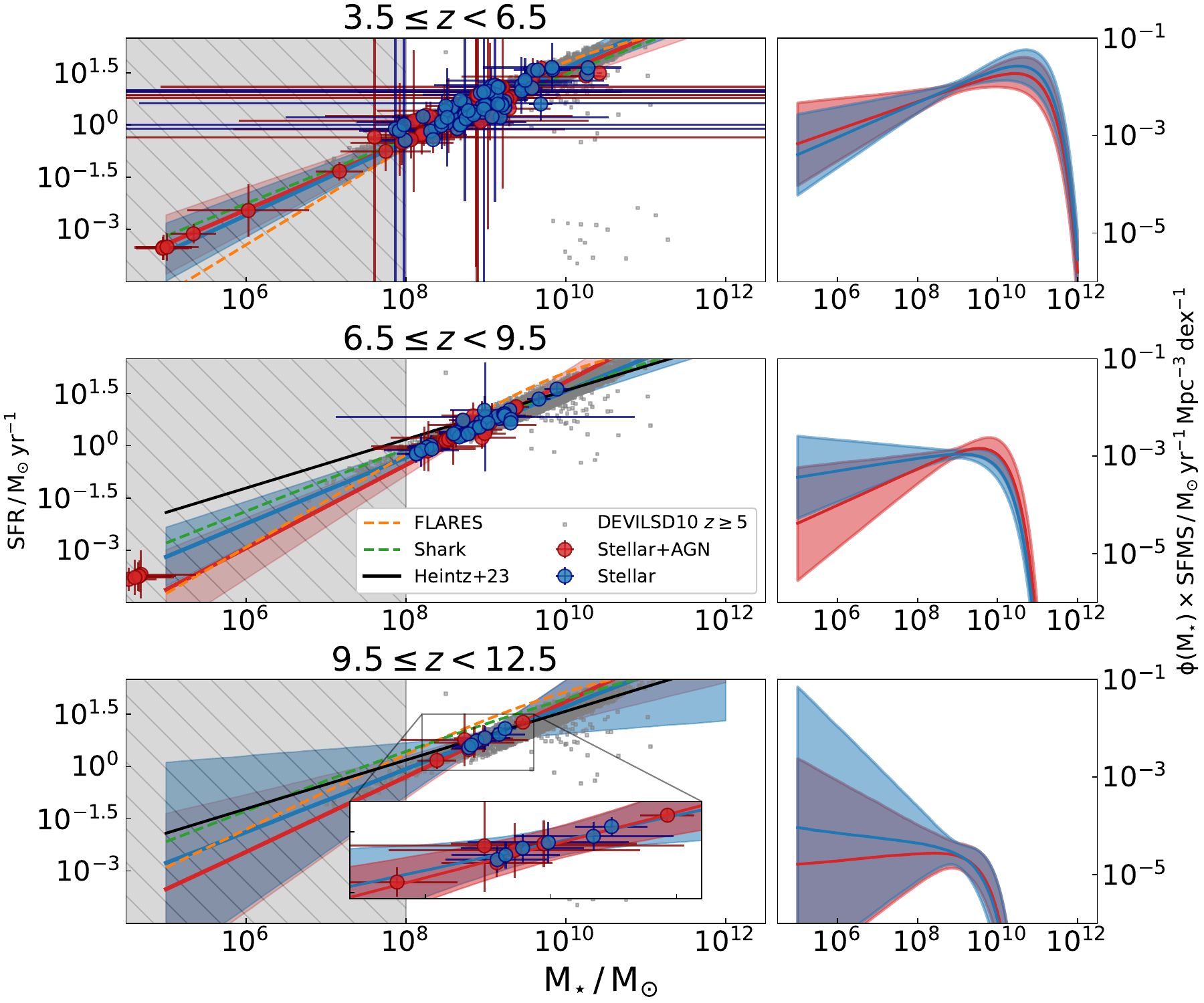}
    \caption{\textit{Left:} SFMS at $z\approx 3.5 \to 12.5$. Blue points and lines show the results from \textsc{Stellar}, while red shows the results from \textsc{Stellar+AGN}. Grey dots are the $z\geq5$ $\mathrm{SFR-M_{\star}}$ from the DEVILS survey \citep[e.g.,][]{thorne_deep_2021}. The grey shaded region shows the exclusion range up to $\mathrm{M_{\star} = 10^{8}M_{\odot}}$ where the stellar masses may not be robust. The inset panel in the highest-redshift bin shows a zoom-in of the data points and the SFMS. \textit{Right:} SFR density distributions per unit stellar mass at $z\approx 3.5 \to 12.5$. The colour scheme is the same as in the left panels. }
    \label{fig:sfms}
\end{figure*}

The left panels of Fig.~\ref{fig:sfms} show the SFR-$\mathrm{M_{\star}}$ of our galaxies, separated into three equal bins of redshift, from $z\approx3.5 - 12.5$. In this figure we show the results computed from both \textsc{Stellar} and \textsc{Stellar+AGN}. In both runs we clearly observe the existence of the SFMS, and so we computed the function $f(\mathrm{M_{\star}})$ as a log-linear function of stellar mass, i.e., $\mathrm{log_{10}(SFR/M_{\odot}\,yr^{-1}) = A\times \mathrm{log_{10}(M_{\star}/M_{\odot})} + B}$. The motivation for this functional form is twofold. First, this function has the minimum degrees of freedom ($dof=2$) to fit the SFMS. Second, previous works have used double power laws, which amount to two linear functions separated by a break mass in log-space ($dof=4$), to fit the SFMS, but have shown that deviation from a single log-linear component only becomes significant at $z<2.6$ \citep[e.g.,][]{thorne_deep_2021}. We used the fully Bayesian fitting tool \textsc{hyper-fit} \citep{robotham_hyper-fit_2015} to fit the SFMS, where Markov Chain Monte Carlo is employed to sample the posterior distributions of the slope and intercept. Because of the limited number of data points in each bin, we first used linear smoothed-splines to first roughly estimate the slope and intercept, and then assumed Gaussian priors about each of those rough estimates of widths 0.5 and 1.0, respectively. The solid lines in Fig.~\ref{fig:sfms} are from the most likely fit parameters, and the shaded regions are the 16th-84th percentile ranges of the posterior distributions, indicating that we are adequately fitting the SFMS.

While not the focus of this Letter, simply the existence of the SFMS at $z \approx 3.5 - 12.5$ is striking. The SFMS is thought to be evidence for self-regulation in galaxies where episodes of gas accretion and condensation are balanced against stellar and/or AGN feedback \citep{tacchella_confinement_2016,davies_galaxy_2019,matthee_origin_2019,katsianis_evolving_2019,davies_deep_2021}. Thus, we interpret the existence of the SFMS to be a result of regulation being established as early as $z\approx 10$ (the median redshift of our highest-redshift bin), when the Universe was only $\approx 500$ Myr old. The orange and green dashed lines in Fig.~\ref{fig:sfms} are predictions of the SFMS at $z\gtrsim 5$ from the \textsc{FLARES} and \textsc{SHARK} simulations \citep[e.g.,][]{dsilva_unveiling_2023,lovell_first_2021,vijayan_first_2021,lagos_shark_2018}. The black line shows a log-linear fit to the $z=7-10$ SFMS from \citet{heintz_fundamental_2022}, who also use \textit{JWST} observations. It should also be mentioned that \citet{nakajima_jwst_2023} also show the existence of the SFMS at $z\gtrsim4$ with \textit{JWST}. The key point is that the existence of the SFMS within the first few hundred Myr since the Big Bang is supported by galaxy formation simulations and independent analyses of \textit{JWST} observations, further encouraging us of our methodology.

We compare our results to the $z\approx5$ SFR-$M_{\star}$ from the DEVILSD10 survey \citep{thorne_deep_2021}, shown with grey points, where we see that our points exhibit similar trends, giving us confidence in the robustness of the SED fits. We do see that some of the \textsc{Stellar+AGN} points in Fig.~\ref{fig:sfms} are hitting a limit that causes their final stellar mass to cluster at $\mathrm{M_{\star}\lesssim 10^{7}M_{\odot}}$. In these cases the SED fits of these galaxies are completely dominated by the AGN component. To ensure that our results are unbiased against potentially erroneous SED fits, we only fitted the SFMS with galaxies of $\mathrm{M_{\star} \geq 10^{8}M_{\odot}}$ as a conservative threshold. 

\subsubsection{SFR distribution function} \label{subsubsect:sfr_distribition}
We drew on previous determinations of the SMF at $z\gtrsim3.5$ to estimate the effective volumes of our spectroscopic sample. Using \textit{Spitzer}/IRAC observations in the CANDELS and Hubble Ultra Deep fields, \citet{stefanon_galaxy_2021} compute $z=6-10$ SMFs as single-component Schechter functions \citep[e.g.,][]{schechter_analytic_1976,weigel_stellar_2016}, and, as of yet, no other $z\approx 10$ observations of the SMF exist besides those. \citet{stefanon_galaxy_2021} only compute the SMF down to $z\approx6$ and so we require a similar SMF to use in our lowest-redshift bin, $z=3.5-6.5$, to be consistent with the results in the two higher-redshift bins.

We first investigated using the $z\approx 5-8$ SMFs from \citet{song_evolution_2016} that are computed from similar \textit{Spitzer}/IRAC observations in the CANDELS and Hubble Ultra Deep fields as \citet{stefanon_galaxy_2021}. We also investigated the $z\approx 5-7$ SMFs from \citet{weaver_cosmos2020_2023} that use the COSMOS2020 catalogue. We compared these two SMFs with the \citet{stefanon_galaxy_2021} SMFs at $z=6-7$, which is the redshift range where the three studies overlap.

The normalisation and the pivot mass of the \citet{weaver_cosmos2020_2023} SMFs, instead of the \citet{song_evolution_2016} SMFs, best agree with \citet{stefanon_galaxy_2021}, as the systematic offset of these fitted parameters between studies is lower and the parameters themselves are better constrained by \citet{weaver_cosmos2020_2023}. The low-mass power-law slope from \citet[][]{song_evolution_2016}, however, agrees better with \citet{stefanon_galaxy_2021}. The low-mass power law slope is fixed by \citet{weaver_cosmos2020_2023}, being $\approx 30$~per cent shallower than the results from \citet{stefanon_galaxy_2021}, and the resulting CSFH is $\approx 0.4$~dex less than that computed with the \citet{song_evolution_2016} SMF. Because of the fixed slope in \citet{weaver_cosmos2020_2023}, we decided to simply use the \citet{song_evolution_2016} SMF to compute the CSFH in our lowest-redshift bin.

While exploring the differences between the underlying SMFs is not the focus of this work, we remark that the adopted SMF will affect the final CSFH. The origin of the difference likely resides in the assumed selection function and resulting effective volumes for the galaxies. These SMFs also used photometric redshifts, and so we are potentially introducing implicit photometric redshift uncertainty into our final CSFH. While it is not ideal to use a mixture of SMFs in different redshift bins, we stress that the aim of this work is to explore the \textit{relative} difference between the \textsc{Stellar+AGN} and \textsc{Stellar} CSFH. As we used the same SMFs for each of these samples, the relative difference does not depend on the exact choice of SMF. Nevertheless, in a follow-up work, we intend to use our own photometric redshifts and selection functions based on joint \textit{HST}+\textit{JWST} photometry to compute the direct SFR density distribution (along with the SMF). In which case, we hope to establish convergence on the true CSFH in the $z\gtrsim3.5$ Universe by means of many independent approaches.

The right panels of Fig.~\ref{fig:sfms} show the SFR density distributions per unit stellar mass. As we wish to highlight the relative effect of the AGN component on the CSFH, we used the same SMFs without an assumed uncertainty for both the \textsc{Stellar} and \textsc{Stellar+AGN} results. Thus, the quantiles show the 16th-84th uncertainty intervals only obtained from the Monte Carlo sampling of the 1000 spline fits of the SFMS. 

There are systematic differences on the final SFR density distribution between the \textsc{Stellar} and \textsc{Stellar+AGN} results, with the \textsc{Stellar} showing higher normalisation at $\mathrm{M_{\star}\approx10^{8} - 10^{10} M_{\odot}}$ and a shallower turnover to lower stellar masses at all redshifts shown. We see that the $9.5 \leq z < 12.5$ bin exhibits diverging SFR density at $\mathrm{M_{\star}\lesssim10^{9}}$, meaning that we can only calculate an upper limit on the final CSFH. We note that the stellar mass density functions from \citet{stefanon_galaxy_2021} are themselves diverging at low stellar masses. Nevertheless, this final SFR density satisfies the integrand of Eq.~\ref{eq:csfh} needed to calculate the CSFH.

\section{Results and Discussion} \label{sec:results}

\subsection{Cosmic star formation history} \label{subsec:csfh}

While we only fitted the SFMS for galaxies with $\mathrm{M_{\star}\geq10^{8}M_{\odot}}$, we extrapolated the log-linear SFMS for the integration of the SFR density function. We integrated the SFR density curves in the domain $10^{5} \mathrm{M_{\odot}} \leq \mathrm{M_{\star}}  < 10^{12} \mathrm{M_{\odot}} $. The bounds of the integration were chosen to encompass a range of galaxies that contribute to the CSFH. The lower limit of $10^{5} \mathrm{M_{\odot}}$ is motivated by stellar mass estimates of Segue 2 \citep{belokurov_discovery_2009,kirby_segue_2013} that is the least massive galaxy discovered, and is likely a tidally stripped dwarf galaxy. The upper limit is motivated by theoretical considerations of the most massive galaxy that can exist in $\Lambda$CDM \citep[e.g.,][]{lovell_extreme_2022}. Note, that we find little enhancement in the CSFH when we instead integrate up to $\mathrm{M_{\star}} = 10^{13} \mathrm{M_{\odot}}$. 

\begin{figure*}
    \centering
    \includegraphics[width=\textwidth]{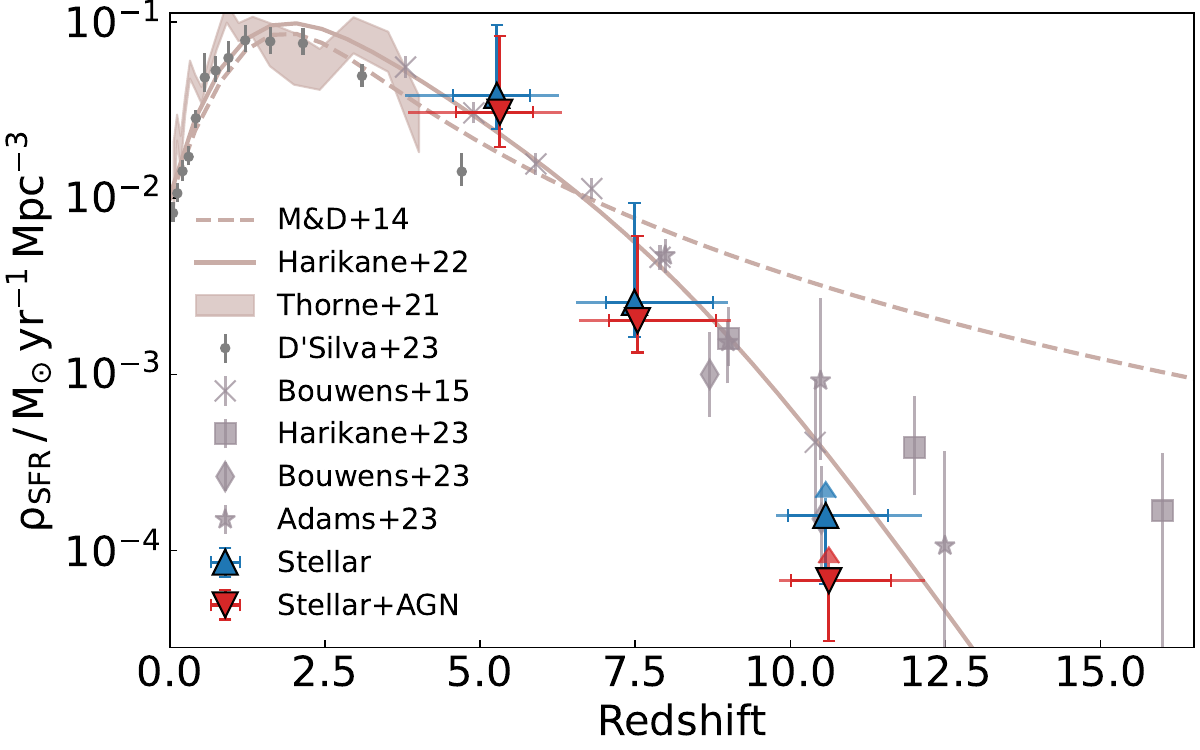}
    \caption{CSFH at $z \approx 0-15$. Blue triangles show the $z\approx 3.5-12.5$ \textsc{Stellar} results, while red inverted triangles show the results for \textsc{Stellar+AGN}. The error bars show the 16th-84th percentile from the SFMS fits in Fig.~\ref{fig:sfms}. The arrows indicate where the results are upper limits. The blue and red points are offset by 0.05 in $z$ for clarity. The brown dashed line shows the \citet{madau_cosmic_2014} fitted CSFH. The brown solid line shows the constant star formation efficiency model of the CSFH from \citet{harikane_goldrush_2022}. The brown shaded region shows the CSFH calculated at $z\approx 0-4$ using the SFMSs and SMFs from \citet{thorne_deep_2021}. The grey points show previous literature results of the CSFH from \citet{dsilva_gamadevils_2023,bouwens_uv_2015, harikane_comprehensive_2023, bouwens_evolution_2023,adams_epochs_2023}.}
    \label{fig:csfh}
\end{figure*}

Fig.~\ref{fig:csfh} shows the resulting CSFH obtained from our sample. As a validation of our method to calculate the CSFH, we used the $z \approx 0-4$  SMFs and SFMSs from \citet{thorne_deep_2021} and compare the final CSFH with the results of \citet{dsilva_gamadevils_2023} in the same redshift range. The results from \citet{dsilva_gamadevils_2023} are direct detections of the CSFH where the full SFR distribution function is integrated. Both of these data sets were built from the same sample of GAMA+DEVILS galaxies and the same \textsc{ProSpect} fits meaning it is correct to compare them. Overall, the two results are consistent, and so we build confidence that this same method, using the SMF and SFMS, can recover the CSFH for our \textit{JWST} sample. 

Moving on, our results at $z\gtrsim3.5$ agree with previous estimates from \citet{bouwens_uv_2015, bouwens_uv_2023, harikane_comprehensive_2023, adams_epochs_2023}. There is a tentative systematic offset between the \textsc{Stellar} and the \textsc{Stellar+AGN} results, ranging from $\approx 0.1$~dex at $z\approx5$ to $\approx 0.4$~dex at $z\approx10.5$, indicating that the AGN component can potentially significantly contaminate the resulting star formation rates in $z>3.5$ galaxies. Estimates of the CSFH at $z>3.5$ generally rely on the UV luminosity function. Indeed, this is the case for the $z\gtrsim 10$ estimates of the CSFH from \citet{harikane_comprehensive_2023, bouwens_evolution_2023, adams_epochs_2023}. But the UV luminosity function is known to be derived from a combination of stellar and AGN UV photons \citep{finkelstein_coevolution_2022}, and so using it to compute the CSFH hinges on a careful delineation between the two processes.

\subsection{AGN growth at $z \gtrsim 10$}
In light of this, we remark that there is a greater offset between the \textsc{Stellar} and \textsc{Stellar+AGN} estimates in the $z\approx 9.5-12.5$ bin than at the lower-redshift bins. Though, we see a higher uncertainty in this highest-redshift bin, on account of our SFR density distributions basically diverging at $\mathrm{M_{\star} \lesssim 10^{8}M_{\odot}}$. Nevertheless, this increasing offset at $z\gtrsim 9.5$ is reminiscent of the offset reported by \citet{harikane_comprehensive_2023} between their CSFH and the constant efficiency models of the CSFH presented in \citet{harikane_goldrush_2022}, which we show by the solid line in Fig.~\ref{fig:csfh}. As such, intuiting these reported offsets to be of common origins, the higher star formation efficiency observed at $z\gtrsim10$ may instead be AGN masquerading as star formation. 

Indeed, \textit{JWST} has confirmed that SMBHs are common at these redshifts \citep[e.g.,][]{larson_ceers_2023,kocevski_hidden_2023,ubler_ga-nifs_2023}, can accrete at super-Eddington rates \citep{schneider_are_2023} and outpace the growth of the host galaxy's stellar mass \citep{maiolino_jades_2023}. Furthermore, \citet{harikane_jwstnirspec_2023} found 10 H$\alpha$ broad-line (FWHM $\approx 1000-6000\,\mathrm{km \, s^{-1}}$) type-1 AGN at $z=3.8-8.9$ from CEERS and GLASS NIRSpec spectra. They suggest that the fraction of galaxies hosting type-1 AGN is higher at $z=4-7$ than it is at $z\approx0$, and that AGN are abundant in the early Universe. We performed a coordinate match between our sample and theirs and found all possible matches. As we were able to find all type-1 AGN identified by \citet{harikane_jwstnirspec_2023} that overlap with our imaging, this is further evidence of the high AGN contribution at these redshifts.

On the contrary, \citet{harikane_comprehensive_2023} concluded that $9/10$ of their $z\approx 12-16$ galaxies cannot possess significant AGN, and thus cannot explain the offset with the constant efficiency model as a result of AGN contamination, on account of their extended morphologies. Indeed, recent results with \textit{JWST} have shown that galaxy disks were already in place at $z\approx 5-8$ \citep{ferreira_jwst_2022,ormerod_epochs_2023}. This implies that the relative weight to the entire morphology of a central point source (AGN) compared to an extended structure (SFR) would be low in these galaxies. This aligns with the low AGN fraction at $z\approx 5$ reported in \citet{dsilva_gamadevils_2023}, where AGN fraction is the ratio of AGN contributed flux between rest-frame $5-20\mu$m to the total bolometric flux of the galaxy. However, this is only a relative assessment of star formation/AGN contribution: a galaxy at high redshift can have a low AGN fraction but a bolometrically luminous AGN. \citet{thorne_deep_2022} suggest that galaxies with a significant AGN contribution have an AGN fraction of $10$~per cent, meaning that significant AGN can even inhabit galaxies where $90$~per cent of the light is constrained to an extended morphological structure. In other words, while extended morphology is a clear discriminant of AGN in cases where the AGN either contributes to all or none of the SED (e.g., AGN fraction $=$ 1 or 0), its significance when we consider the union of star formation and AGN is less so. Thus, the spatial light distributions need not necessarily be centrally concentrated point sources for $z \gtrsim 9.5$ galaxies hosting powerful AGN because extended star formation can be just as active during the lead up to the peak of CSFH. 

\subsection{AGN effect on host galaxies}
On the basis that our galaxies can in fact host powerful AGN, we further investigate the effect of the AGN on the properties of the host galaxies.

\begin{figure*}[h]
    \centering
    \includegraphics[width=\textwidth]{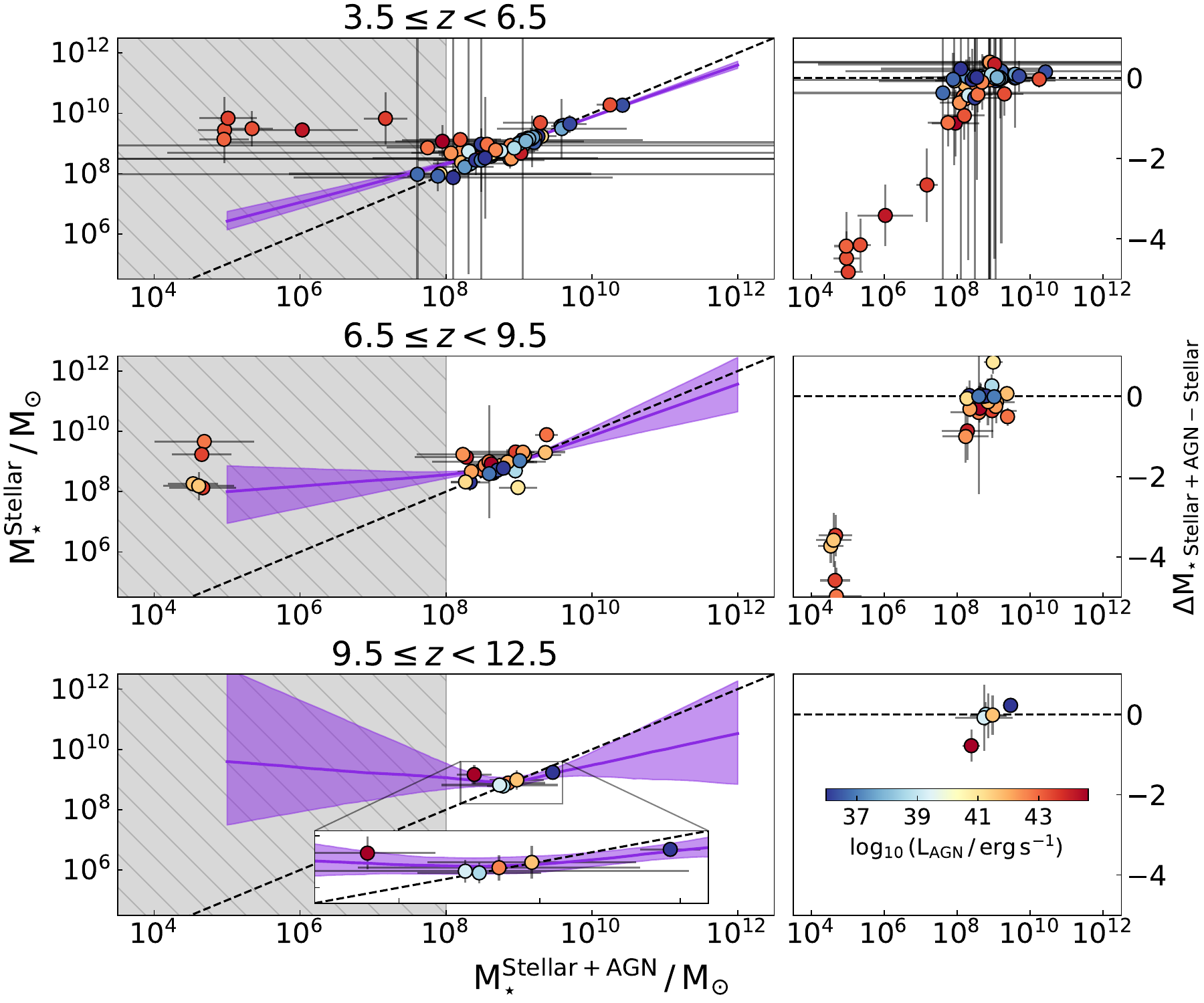}
    \caption{\textit{Left:} \textsc{Stellar} $\mathrm{M_{\star}}$ against \textsc{Stellar+AGN} $\mathrm{M_{\star}}$ at $z\approx 3.5-12.5$. The colour of the points corresponds to the AGN bolometric luminosity determined by \textsc{ProSpect}, where blue is weaker AGN and red is stronger. The purple line and shading show the third-order spline fit. The grey shaded region shows exclusion range up to $\mathrm{M_{\star} = 10^{8}M_{\odot}}$ where the stellar masses may not be robust. The dashed line shows the one-to-one relation. The inset panel in the highest-redshift bin shows a zoom-in of the data points and fitted relation. \textit{Right:} $\mathrm{\Delta M_{\star}=}$ \textsc{Stellar+AGN} $-$ \textsc{Stellar}. The colour scheme is the same as the left panels. The horizontal dashed line shows $\mathrm{\Delta M_{\star}=0}$.}
    \label{fig:deltamstar}
\end{figure*}

Fig.~\ref{fig:deltamstar} shows the difference in stellar mass between the \textsc{Stellar} and \textsc{Stellar+AGN} SED fits to show what effect accounting for an AGN component has on the fitting. Galaxies with $\mathrm{M_{\star}} \gtrsim 10^{8} \mathrm{M_{\odot}}$ computed from both \textsc{ProSpect} runs have similar derived stellar masses. To further explore the trends, we fitted the relations with third-order smoothing splines, where the three degrees of freedom allow us to see any inflection beyond a log-linear correlation. For the same reasons that we expressed in fitting the SFMS in Section~\ref{subsubsect:fitting_sfms}, we only performed fitting for galaxies with $\mathrm{M_{\star} \geq 10^{8} \mathrm{M_{\odot}}}$ (from both \textsc{ProSpect} runs). Overall, we see that in the $\mathrm{M_{\star} \gtrsim 10^{9} M_{\odot}}$ quadrant at all redshifts, the results agree within uncertainties. There is a tentative inclination for \textsc{Stellar+AGN} to produce more massive galaxies, but this is observed only well into the extrapolated fit and so we cannot make robust conclusions here. 

In the $\mathrm{M_{\star} \lesssim 10^{9} M_{\odot}}$ quadrant we see that that relation turns up showing that \textsc{Stellar+AGN} produces systematically lower stellar masses than \textsc{Stellar} for these galaxies. This is consistent with the study of the bright ($\mathrm{M_{UV}=-21.6\,AB\, mag}$), $z=10.6$ galaxy, GN-z11, whose nuclear morphological component contributes $2/3$ of the total UV luminosity but only $1/5$ of the stellar mass \citep{tacchella_jades_2023}. Thus, for galaxies that host powerful AGN, which are convincingly abundant at $z\gtrsim 5$, the UV luminosity predominately originates from accelerating charged particles in the accretion disk around the SMBH and not from star formation, where only the latter process contributes to stellar mass.

Turning our attention to the exact difference in stellar mass between runs, accounting for an AGN, incredibly, can shift the final stellar mass by up to $\approx 4$~dex. For these galaxies, the SED is basically fit entirely by the AGN component as evidenced by the greatest offset for galaxies containing the most luminous AGN. Though, we repeat that this is highly suggestive that the \textsc{Stellar+AGN} fitting is hitting a limit in stellar mass and SFR. The limit is due to the mass-to-light ratio approaching the minimum allowable value from \textsc{ProSpect's} set up of \citet{bruzual_stellar_2003} stellar population models when the SED is modelled overwhelmingly by an AGN component. As a consequence of this limit these fits may not be robust. Nevertheless, such a significant difference between \textsc{ProSpect} runs indicates to us that AGN is an important factor in interpreting the SEDs of $z \approx 3.5-12.5$ galaxies.

Tantalisingly, these results may offer an avenue to explain massive, quenched galaxies at $z\gtrsim 3$ whose formation times are challenging to reconcile with the state-of-the-art galaxy formation models \citep[e.g.,][]{lovell_extreme_2022, carnall_surprising_2023, glazebrook_extraordinarily_2023}. Sure enough, recent theoretical works have shown that quiescent galaxies can exist at these redshifts in quantities that agree with observations and that among the main drivers of star formation suppression is feedback from the AGN \citep{lovell_first_2023,lagos_quenching_2023}. In essence, alternative prescriptions that account for the AGN, both from the perspective of SED modelling in the observations and subgrid physics implementation in the simulations, can explain the existence of these massive galaxies.

\begin{figure*}
    \centering
    \includegraphics[width=\textwidth]{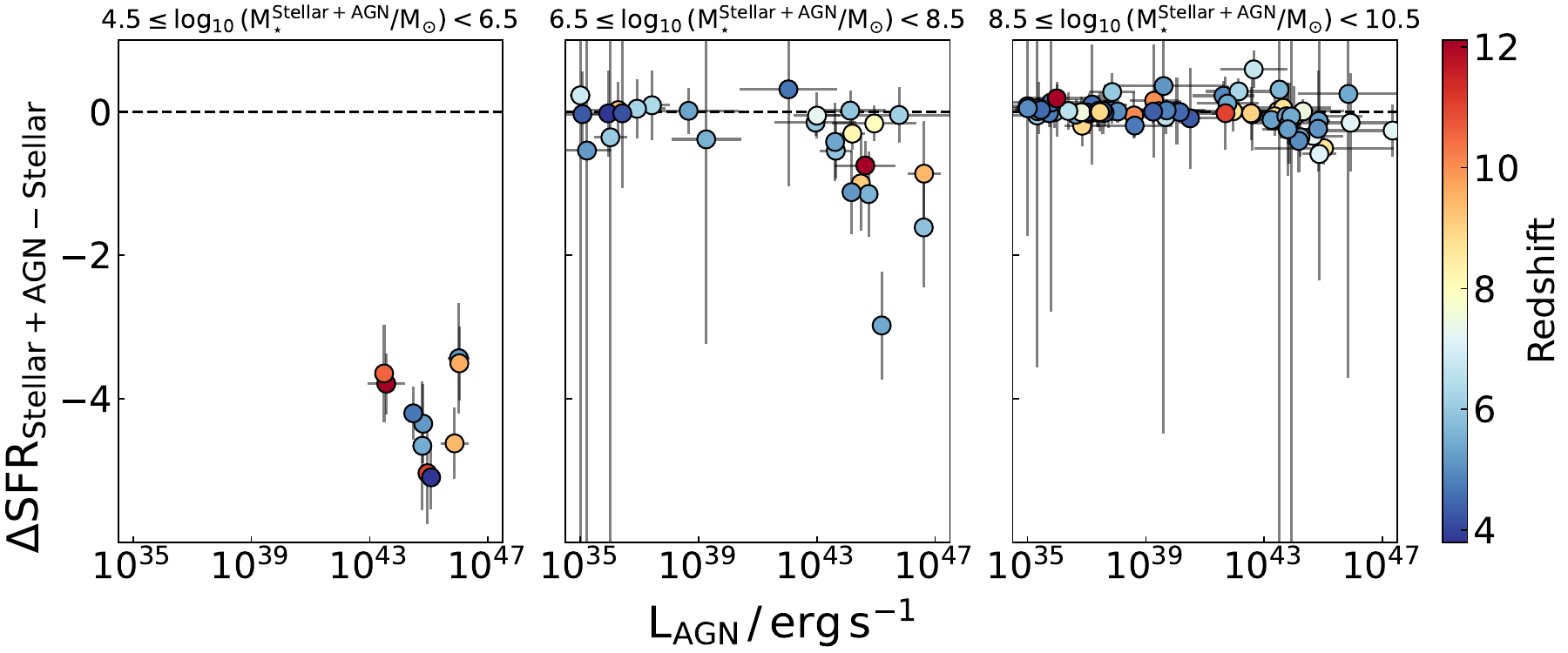}
    \caption{$\mathrm{\Delta SFR}=$ \textsc{Stellar+AGN} $-$ \textsc{Stellar} as a function of AGN bolometric luminosity. We show the results binned in $2$~dex intervals of \textsc{Stellar+AGN} stellar mass from $\mathrm{M_{\star}\approx 10^{4.5}M_{\odot} \to 10^{10.5}M_{\odot}}$. The colour of points indicates the redshift, where blue corresponds to low redshift and red to high. The horizontal dashed line shows $\mathrm{\Delta SFR}=0$.}
    \label{fig:deltasfr}
\end{figure*}

Offsets in stellar mass between the two \textsc{ProSpect} runs is the cumulative effect of the difference in the underlying SFR. In Fig.~\ref{fig:deltasfr} we show the difference in SFR between the two runs as a function of AGN bolometric luminosity. A clear trend is observed where galaxies that \textsc{Stellar+AGN} has determined to host powerful, $\mathrm{L_{AGN}} \approx 10^{43} \, \mathrm{erg \, s^{-1}}$, AGN experience as much as a $\approx 4$~dex difference in SFR between runs. We also see that these galaxies have tiny stellar masses, $\mathrm{M_{\star} \lesssim 10^{6.5} M_{\odot}}$, when determined from the \textsc{Stellar+AGN} fits. This difference in SFR toward higher AGN bolometric luminosity is the genesis of the difference in our CSFH, seen in Fig.~\ref{fig:csfh}. 

What does this mean in terms of the physics of star formation and AGN activity at $z\gtrsim 3.5$? Well, we interpret the offset in CSFH between the two \textsc{ProSpect} runs to mean that star formation does not overwhelm AGN activity (with respect to the relative contributions to the bolometric luminosity of the galaxy), especially at $z\gtrsim 9.5$. Indeed, as the gas supply is abundant at high redshift we can intuit that the same gas reservoir fuelling star formation is also feeding the AGN. This agrees with the picture outlined in \citet{dsilva_gamadevils_2023} who showed that the common depletion of the gas supply from $z\approx 2 \to 0$ produces a coeval decline in both the star formation and AGN activity.

Whether this coevolution persists at $z\gtrsim 2$ remains elusive. Recent results with \textit{JWST} tend to indicate that SMBHs are overmassive compared to their host galaxy's stellar mass \citep[e.g.,][]{maiolino_jades_2023} on account of super-Eddington accretion onto the SMBH \citep[e.g.,][]{schneider_are_2023}, pointing to an unequal growth of stellar mass compared to black hole mass. Interestingly, these objects may follow a formation mechanism where massive, metal-poor gas clouds directly collapse into SMBHs whose accretion disk dominates the contribution to the SED compared to the stellar component \citep[e.g.,][]{bromm_formation_2003,agarwal_unravelling_2013,kroupa_very_2020}.

To further unveil the intricacies of star formation and AGN activity at $z\gtrsim 2-5$, in a future work, D'Silva et. al (in preparation), we intend to use large photometric samples of \textit{JWST} high-redshift candidates from CEERS \citep[Program: 1345, PI: S. Finkelstein;][]{finkelstein_ceers_2023,bagley_ceers_2023}, GLASS \citep[Program: 1324, PI: T. Treu;][]{treu_glass-jwst_2022}, UNCOVER \citep[Program: 2561, PI: I. Labbe;][]{bezanson_jwst_2022}(PID 2561), NGDEEP \citep[Program: 2079, PI: S. Finkelstein;][]{bagley_next_2023}, COSMOS WEB \citep[Program: 1727, PI: J. Kartaltepe;][]{casey_cosmos-web_2023} and PEARLS \citep[Programs 1176 and 2738, PI: R. Windhorst;][]{windhorst_webbs_2022}. The work presented in this Letter is thus a stepping stone and anchor point to address the coevolution of star formation and AGN within the first galaxies. 

\section{Conclusion} \label{sec:conclusion}
Central to the science aspirations of \textit{JWST} is shedding light on the nature of stellar mass assembly and the growth of SMBHs within the first few billion years after the Big Bang. \textit{JWST} has so far generated more questions than it has answered about the $z\approx 3.5-12.5$ frontier. A key discussion freshly sparked by \textit{JWST} is whether star formation is more efficient than we previously expected. In this work, we performed SED fitting on 109 spectroscopically confirmed $z\approx 3.5-12.5$ to see if AGN masquerading as star formation can alleviate this tension. Our main findings are:

\begin{itemize}
    \item The CSFH is tentatively as much as $\approx 0.4$~dex lower when computed using SFRs derived from the \textsc{Stellar+AGN} fits that simultaneously model the stars and the AGN. Thus, a hidden AGN component can potentially explain the excess of CSFH compared to constant star formation efficiency models. 

    \item \textsc{Stellar+AGN} fits produce as much as $\approx 4$~dex lower stellar masses and SFRs compared to \textsc{Stellar}. This is essentially the origin of the offset between the CSFH. Thus, a hidden AGN component may be a promising avenue to explain massive quenched galaxies at $z\gtrsim3$ that are seemingly at odds with current theories of stellar mass assembly. However, the large difference between \textsc{ProSpect} runs is suggestive that we are hitting a limit in SFR and stellar mass in the SED fitting. Nevertheless, this highlights that AGN may be an important factor in interpreting the SEDs of $z\approx 3.5-12.5$ galaxies.
\end{itemize}

 We hope that this work highlights the importance of considering the union of star formation and AGN activity within the first galaxies to ever exist. 

\section{Data availability}
Analysis scripts used in this work are available at \url{https://github.com/JordanDSilva/project2_jwst_csfh} and here \dataset[10.5281/zenodo.10148456]{http://dx.doi.org/10.5281/zenodo.10148456}. SED fits, photometry and imaging used in this work will be available upon reasonable request to the corresponding author. \textit{JWST} data used in this work can be found at \dataset[10.17909/t317-yp54]{http://dx.doi.org/10.17909/t317-yp54} and \dataset[10.17909/akh9-5057]{http://dx.doi.org/10.17909/akh9-5057}.

\begin{acknowledgments}
We thank the anonymous referee/s for reviewing and improving the quality of this work. J.C.J.D is supported by the Australian Government Research Training Program (RTP) Scholarship. S.P.D acknowledges funding by the Australian Research Council (ARC) Laureate Fellowship scheme (FL220100191). A.S.G.R acknowledges funding by the Australian Research Council (ARC) Future Fellowship scheme (FT200100375). This research was supported by the Australian Research Council Centre of Excellence for All Sky Astrophysics in 3 Dimensions (ASTRO 3D), through project number CE170100013.  This work is based on observations made with the NASA/ESA Hubble Space Telescope (\textit{HST}) and NASA/ESA/CSA James Webb Space Telescope (\textit{JWST}) obtained from the Mikulski Archive for Space Telescopes (MAST) at the Space Telescope Science Institute (STScI), which is operated by the Association of Universities for Research in Astronomy, Inc., under NASA contract NAS 5-03127 for JWST, and NAS 5–26555 for HST. The observations used in this work are associated with JWST programs 1345 and 1324. We acknowledge all those involved in the development of the \textit{JWST} and the execution of these large observing programs. 
\end{acknowledgments}

%

\vspace{5mm}








\bibliography{ref}{}

\begin{thebibliography}{}
\expandafter\ifx\csname natexlab\endcsname\relax\def\natexlab#1{#1}\fi
\providecommand{\url}[1]{\href{#1}{#1}}
\providecommand{\dodoi}[1]{doi:~\href{http://doi.org/#1}{\nolinkurl{#1}}}
\providecommand{\doeprint}[1]{\href{http://ascl.net/#1}{\nolinkurl{http://ascl.net/#1}}}
\providecommand{\doarXiv}[1]{\href{https://arxiv.org/abs/#1}{\nolinkurl{https://arxiv.org/abs/#1}}}

\bibitem[{{Adams} {et~al.}(2023){Adams}, {Conselice}, {Ferreira}, {Austin},
  {Trussler}, {Juod{\v{z}}balis}, {Wilkins}, {Caruana}, {Dayal}, {Verma}, \&
  {Vijayan}}]{adams_discovery_2023}
{Adams}, N.~J., {Conselice}, C.~J., {Ferreira}, L., {et~al.} 2023, \mnras, 518,
  4755, \dodoi{10.1093/mnras/stac3347}

\bibitem[{Adams {et~al.}(2023)Adams, Conselice, Austin, Harvey, Ferreira,
  Trussler, Juodzbalis, Li, Windhorst, Cohen, Jansen, Summers, Tompkins,
  Driver, Robotham, D'Silva, Yan, Coe, Frye, Grogin, Koekemoer, Marshall,
  Pirzkal, Ryan, Maksym, Rutkowski, Willmer, Hammel, Nonino, Bhatawdekar,
  Wilkins, Willner, Bradley, Broadhurst, Cheng, Dole, Hathi, \&
  Zitrin}]{adams_epochs_2023}
Adams, N.~J., Conselice, C.~J., Austin, D., {et~al.} 2023, {EPOCHS} {Paper}
  {II}: {The} {Ultraviolet} {Luminosity} {Function} from
  \$7.5{\textless}z{\textless}13.5\$ using 110 square arcminutes of deep,
  blank-field data from the {PEARLS} {Survey} and {Public} {Science}
  {Programmes},  arXiv.
\newblock \url{http://arxiv.org/abs/2304.13721}

\bibitem[{Agarwal {et~al.}(2013)Agarwal, Davis, Khochfar, Natarajan, \&
  Dunlop}]{agarwal_unravelling_2013}
Agarwal, B., Davis, A.~J., Khochfar, S., Natarajan, P., \& Dunlop, J.~S. 2013,
  Monthly Notices of the Royal Astronomical Society, 432, 3438,
  \dodoi{10.1093/mnras/stt696}

\bibitem[{{Arrabal Haro} {et~al.}(2023){Arrabal Haro}, {Dickinson},
  {Finkelstein}, {Fujimoto}, {Fern{\'a}ndez}, {Kartaltepe}, {Jung}, {Cole},
  {Burgarella}, {Chworowsky}, {Hutchison}, {Morales}, {Papovich}, {Simons},
  {Amor{\'\i}n}, {Backhaus}, {Bagley}, {Bisigello}, {Calabr{\`o}},
  {Castellano}, {Cleri}, {Dav{\'e}}, {Dekel}, {Ferguson}, {Fontana}, {Gawiser},
  {Giavalisco}, {Harish}, {Hathi}, {Hirschmann}, {Holwerda}, {Huertas-Company},
  {Koekemoer}, {Larson}, {Lucas}, {Mobasher}, {P{\'e}rez-Gonz{\'a}lez},
  {Pirzkal}, {Rose}, {Santini}, {Trump}, {de la Vega}, {Wang}, {Weiner},
  {Wilkins}, {Yang}, {Yung}, \& {Zavala}}]{haro_spectroscopic_2023}
{Arrabal Haro}, P., {Dickinson}, M., {Finkelstein}, S.~L., {et~al.} 2023,
  \apjl, 951, L22, \dodoi{10.3847/2041-8213/acdd54}

\bibitem[{Bagley {et~al.}(2023{\natexlab{a}})Bagley, Finkelstein, Koekemoer,
  Ferguson, Haro, Dickinson, Kartaltepe, Papovich, Pérez-González, Pirzkal,
  Somerville, Willmer, Yang, Yung, Fontana, Grazian, Grogin, Hirschmann,
  Kewley, Kirkpatrick, Kocevski, Lotz, Medrano, Morales, Pentericci,
  Ravindranath, Trump, Wilkins, Calabrò, Cooper, Costantin, de~la Vega,
  Hutchison, Lucas, McGrath, Wang, \& Wuyts}]{bagley_ceers_2023}
Bagley, M.~B., Finkelstein, S.~L., Koekemoer, A.~M., {et~al.}
  2023{\natexlab{a}}, The Astrophysical Journal Letters, 946, L12,
  \dodoi{10.3847/2041-8213/acbb08}

\bibitem[{Bagley {et~al.}(2023{\natexlab{b}})Bagley, Pirzkal, Finkelstein,
  Papovich, Berg, Lotz, Leung, Ferguson, Koekemoer, Dickinson, Kartaltepe,
  Kocevski, Somerville, Yung, Backhaus, Casey, Castellano, Ortiz, Chworowsky,
  Cox, Davé, Davis, Estrada-Carpenter, Fontana, Fujimoto, Gardner, Giavalisco,
  Grazian, Grogin, Hathi, Hutchison, Jaskot, Jung, Kewley, Kirkpatrick, Larson,
  Matharu, Natarajan, Pentericci, Pérez-González, Ravindranath, Rothberg,
  Ryan, Shen, Simons, Snyder, Trump, \& Wilkins}]{bagley_next_2023}
Bagley, M.~B., Pirzkal, N., Finkelstein, S.~L., {et~al.} 2023{\natexlab{b}},
  The {Next} {Generation} {Deep} {Extragalactic} {Exploratory} {Public}
  ({NGDEEP}) {Survey},  arXiv, \dodoi{10.48550/arXiv.2302.05466}

\bibitem[{Bakx {et~al.}(2023)Bakx, Zavala, Mitsuhashi, Treu, Fontana, Tadaki,
  Casey, Castellano, Glazebrook, Hagimoto, Ikeda, Jones, Leethochawalit, Mason,
  Morishita, Nanayakkara, Pentericci, Roberts-Borsani, Santini, Serjeant,
  Tamura, Trenti, \& Vanzella}]{bakx_deep_2023}
Bakx, T. J. L.~C., Zavala, J.~A., Mitsuhashi, I., {et~al.} 2023, Monthly
  Notices of the Royal Astronomical Society, 519, 5076,
  \dodoi{10.1093/mnras/stac3723}

\bibitem[{Bellstedt {et~al.}(2020{\natexlab{a}})Bellstedt, Driver, Robotham,
  Davies, Bogue, Cook, Hashemizadeh, Koushan, Taylor, Thorne, Turner, \&
  Wright}]{bellstedt_galaxy_2020-1}
Bellstedt, S., Driver, S.~P., Robotham, A. S.~G., {et~al.} 2020{\natexlab{a}},
  Monthly Notices of the Royal Astronomical Society, 496, 3235,
  \dodoi{10.1093/mnras/staa1466}

\bibitem[{Bellstedt {et~al.}(2020{\natexlab{b}})Bellstedt, Robotham, Driver,
  Thorne, Davies, Lagos, Stevens, Taylor, Baldry, Moffett, Hopkins, \&
  Phillipps}]{bellstedt_galaxy_2020}
Bellstedt, S., Robotham, A. S.~G., Driver, S.~P., {et~al.} 2020{\natexlab{b}},
  Monthly Notices of the Royal Astronomical Society, 498, 5581,
  \dodoi{10.1093/mnras/staa2620}

\bibitem[{Belokurov {et~al.}(2009)Belokurov, Walker, Evans, Gilmore, Irwin,
  Mateo, Mayer, Olszewski, Bechtold, \& Pickering}]{belokurov_discovery_2009}
Belokurov, V., Walker, M.~G., Evans, N.~W., {et~al.} 2009, Monthly Notices of
  the Royal Astronomical Society, 397, 1748,
  \dodoi{10.1111/j.1365-2966.2009.15106.x}

\bibitem[{Bertin \& Arnouts(1996)}]{bertin_sextractor_1996}
Bertin, E., \& Arnouts, S. 1996, Astronomy and Astrophysics Supplement, v.117,
  p.393-404, 117, 393, \dodoi{10.1051/aas:1996164}

\bibitem[{Bezanson {et~al.}(2022)Bezanson, Labbe, Whitaker, Leja, Price, Franx,
  Brammer, Marchesini, Zitrin, Wang, Weaver, Furtak, Atek, Coe, Cutler, Dayal,
  van Dokkum, Feldmann, Schreiber, Fujimoto, Geha, Glazebrook, de~Graaff,
  Greene, Juneau, Kassin, Kriek, Khullar, Maseda, Mowla, Muzzin, Nanayakkara,
  Nelson, Oesch, Pacifici, Pan, Papovich, Setton, Shapley, Smit, Stefanon,
  Taylor, \& Williams}]{bezanson_jwst_2022}
Bezanson, R., Labbe, I., Whitaker, K.~E., {et~al.} 2022, The {JWST} {UNCOVER}
  {Treasury} survey: {Ultradeep} {NIRSpec} and {NIRCam} {ObserVations} before
  the {Epoch} of {Reionization},  arXiv.
\newblock \url{http://arxiv.org/abs/2212.04026}

\bibitem[{Bouwens {et~al.}(2023{\natexlab{a}})Bouwens, Illingworth, Oesch,
  Stefanon, Naidu, van Leeuwen, \& Magee}]{bouwens_uv_2023}
Bouwens, R., Illingworth, G., Oesch, P., {et~al.} 2023{\natexlab{a}}, Monthly
  Notices of the Royal Astronomical Society, 523, 1009,
  \dodoi{10.1093/mnras/stad1014}

\bibitem[{Bouwens {et~al.}(2015)Bouwens, Illingworth, Oesch, Trenti, Labbé,
  Bradley, Carollo, van Dokkum, Gonzalez, Holwerda, Franx, Spitler, Smit, \&
  Magee}]{bouwens_uv_2015}
Bouwens, R.~J., Illingworth, G.~D., Oesch, P.~A., {et~al.} 2015, The
  Astrophysical Journal, 803, 34, \dodoi{10.1088/0004-637X/803/1/34}

\bibitem[{Bouwens {et~al.}(2023{\natexlab{b}})Bouwens, Stefanon, Brammer,
  Oesch, Herard-Demanche, Illingworth, Matthee, Naidu, van Dokkum, \& van
  Leeuwen}]{bouwens_evolution_2023}
Bouwens, R.~J., Stefanon, M., Brammer, G., {et~al.} 2023{\natexlab{b}}, Monthly
  Notices of the Royal Astronomical Society, 523, 1036,
  \dodoi{10.1093/mnras/stad1145}

\bibitem[{Bromm \& Loeb(2003)}]{bromm_formation_2003}
Bromm, V., \& Loeb, A. 2003, The Astrophysical Journal, 596, 34,
  \dodoi{10.1086/377529}

\bibitem[{Bruzual \& Charlot(2003)}]{bruzual_stellar_2003}
Bruzual, G., \& Charlot, S. 2003, Monthly Notices of the Royal Astronomical
  Society, 344, 1000, \dodoi{10.1046/j.1365-8711.2003.06897.x}

\bibitem[{{Bunker} {et~al.}(2023){Bunker}, {Saxena}, {Cameron}, {Willott},
  {Curtis-Lake}, {Jakobsen}, {Carniani}, {Smit}, {Maiolino}, {Witstok},
  {Curti}, {D'Eugenio}, {Jones}, {Ferruit}, {Arribas}, {Charlot}, {Chevallard},
  {Giardino}, {de Graaff}, {Looser}, {L{\"u}tzgendorf}, {Maseda}, {Rawle},
  {Rix}, {Del Pino}, {Alberts}, {Egami}, {Eisenstein}, {Endsley}, {Hainline},
  {Hausen}, {Johnson}, {Rieke}, {Rieke}, {Robertson}, {Shivaei}, {Stark},
  {Sun}, {Tacchella}, {Tang}, {Williams}, {Willmer}, {Baker}, {Baum},
  {Bhatawdekar}, {Bowler}, {Boyett}, {Chen}, {Circosta}, {Helton}, {Ji},
  {Kumari}, {Lyu}, {Nelson}, {Parlanti}, {Perna}, {Sandles}, {Scholtz},
  {Suess}, {Topping}, {{\"U}bler}, {Wallace}, \& {Whitler}}]{bunker_jades_2023}
{Bunker}, A.~J., {Saxena}, A., {Cameron}, A.~J., {et~al.} 2023, \aap, 677, A88,
  \dodoi{10.1051/0004-6361/202346159}

\bibitem[{Bushouse {et~al.}(2023{\natexlab{a}})Bushouse, Eisenhamer, Dencheva,
  Davies, Greenfield, Morrison, Hodge, Simon, Grumm, Droettboom, Slavich,
  Sosey, Pauly, Miller, Jedrzejewski, Hack, Davis, Crawford, Law, Gordon,
  Regan, Cara, MacDonald, Bradley, Shanahan, Jamieson, Teodoro, \&
  Williams}]{bushouse_jwst_1102_2023}
Bushouse, H., Eisenhamer, J., Dencheva, N., {et~al.} 2023{\natexlab{a}}, {JWST}
  {Calibration} {Pipeline},  Zenodo, \dodoi{10.5281/zenodo.7829329}

\bibitem[{Bushouse {et~al.}(2023{\natexlab{b}})Bushouse, Eisenhamer, Dencheva,
  Davies, Greenfield, Morrison, Hodge, Simon, Grumm, Droettboom, Slavich,
  Sosey, Pauly, Miller, Jedrzejewski, Hack, Davis, Crawford, Law, Gordon,
  Regan, Cara, MacDonald, Bradley, Shanahan, Jamieson, Teodoro, \&
  Williams}]{bushouse_jwst_1112_2023}
---. 2023{\natexlab{b}}, {JWST} {Calibration} {Pipeline},  Zenodo,
  \dodoi{10.5281/zenodo.8140011}

\bibitem[{Cardoso {et~al.}(2017)Cardoso, Gomes, \&
  Papaderos}]{cardoso_impact_2017}
Cardoso, L. S.~M., Gomes, J.~M., \& Papaderos, P. 2017, Astronomy \&
  Astrophysics, 604, A99, \dodoi{10.1051/0004-6361/201630378}

\bibitem[{Carnall {et~al.}(2023)Carnall, McLeod, McLure, Dunlop, Begley,
  Cullen, Donnan, Hamadouche, Jewell, Jones, Pollock, \&
  Wild}]{carnall_surprising_2023}
Carnall, A.~C., McLeod, D.~J., McLure, R.~J., {et~al.} 2023, Monthly Notices of
  the Royal Astronomical Society, 520, 3974, \dodoi{10.1093/mnras/stad369}

\bibitem[{Casey {et~al.}(2023)Casey, Akins, Shuntov, Ilbert, Paquereau, Franco,
  Hayward, Finkelstein, Boylan-Kolchin, Robertson, Allen, Brinch, Cooper, Ding,
  Drakos, Faisst, Fujimoto, Gillman, Harish, Hirschmann, Jin, Kartaltepe,
  Koekemoer, Kokorev, Liu, Long, Magdis, Maraston, Martin, McCracken, McKinney,
  Mobasher, Rhodes, Rich, Sanders, Silverman, Toft, Vijayan, Weaver, Wilkins,
  Yang, \& Zavala}]{casey_cosmos-web_2023}
Casey, C.~M., Akins, H.~B., Shuntov, M., {et~al.} 2023, {COSMOS}-{Web}:
  {Intrinsically} {Luminous} z\${\textbackslash}gtrsim\$10 {Galaxy}
  {Candidates} {Test} {Early} {Stellar} {Mass} {Assembly},  arXiv.
\newblock \url{http://arxiv.org/abs/2308.10932}

\bibitem[{Chabrier(2003)}]{chabrier_galactic_2003}
Chabrier, G. 2003, Publications of the Astronomical Society of the Pacific,
  115, 763, \dodoi{10.1086/376392}

\bibitem[{Charlot \& Fall(2000)}]{charlot_simple_2000}
Charlot, S., \& Fall, S.~M. 2000, The Astrophysical Journal, 539, 718,
  \dodoi{10.1086/309250}

\bibitem[{Conroy(2013)}]{conroy_modeling_2013}
Conroy, C. 2013, Annual Review of Astronomy and Astrophysics, 51, 393,
  \dodoi{10.1146/annurev-astro-082812-141017}

\bibitem[{Curtis-Lake {et~al.}(2023)Curtis-Lake, Carniani, Cameron, Charlot,
  Jakobsen, Maiolino, Bunker, Witstok, Smit, Chevallard, Willott, Ferruit,
  Arribas, Bonaventura, Curti, D'Eugenio, Franx, Giardino, Looser,
  Lützgendorf, Maseda, Rawle, Rix, Rodríguez~del Pino, Übler, Sirianni,
  Dressler, Egami, Eisenstein, Endsley, Hainline, Hausen, Johnson, Rieke,
  Robertson, Shivaei, Stark, Tacchella, Williams, Willmer, Bhatawdekar, Bowler,
  Boyett, Chen, de~Graaff, Helton, Hviding, Jones, Kumari, Lyu, Nelson, Perna,
  Sandles, Saxena, Suess, Sun, Topping, Wallace, \&
  Whitler}]{curtis-lake_spectroscopic_2023}
Curtis-Lake, E., Carniani, S., Cameron, A., {et~al.} 2023, Nature Astronomy, 7,
  622, \dodoi{10.1038/s41550-023-01918-w}

\bibitem[{Dale {et~al.}(2014)Dale, Helou, Magdis, Armus, Díaz-Santos, \&
  Shi}]{dale_two-parameter_2014}
Dale, D.~A., Helou, G., Magdis, G.~E., {et~al.} 2014, The Astrophysical
  Journal, 784, 83, \dodoi{10.1088/0004-637X/784/1/83}

\bibitem[{Davies {et~al.}(2016)Davies, Driver, Robotham, Grootes, Popescu,
  Tuffs, Hopkins, Alpaslan, Andrews, Bland-Hawthorn, Bremer, Brough, Brown,
  Cluver, Croom, da~Cunha, Dunne, Lara-López, Liske, Loveday, Moffett, Owers,
  Phillipps, Sansom, Taylor, Michalowski, Ibar, Smith, \&
  Bourne}]{davies_gamah-atlas_2016}
Davies, L. J.~M., Driver, S.~P., Robotham, A. S.~G., {et~al.} 2016, Monthly
  Notices of the Royal Astronomical Society, 461, 458,
  \dodoi{10.1093/mnras/stw1342}

\bibitem[{Davies {et~al.}(2019)Davies, Lagos, Katsianis, Robotham, Cortese,
  Driver, Bremer, Brown, Brough, Cluver, Grootes, Holwerda, Owers, \&
  Phillipps}]{davies_galaxy_2019}
Davies, L. J.~M., Lagos, C. d.~P., Katsianis, A., {et~al.} 2019, Monthly
  Notices of the Royal Astronomical Society, 483, 1881,
  \dodoi{10.1093/mnras/sty2957}

\bibitem[{Davies {et~al.}(2021)Davies, Thorne, Robotham, Bellstedt, Driver,
  Adams, Bilicki, Bowler, Bravo, Cortese, Foster, Grootes, Häußler,
  Hashemizadeh, Holwerda, Hurley, Jarvis, Lidman, Maddox, Meyer, Paolillo,
  Phillipps, Radovich, Siudek, Vaccari, \& Windhorst}]{davies_deep_2021}
Davies, L. J.~M., Thorne, J.~E., Robotham, A. S.~G., {et~al.} 2021, Monthly
  Notices of the Royal Astronomical Society, 506, 256,
  \dodoi{10.1093/mnras/stab1601}

\bibitem[{Davies {et~al.}(2022)Davies, Thorne, Bellstedt, Bravo, Robotham,
  Driver, Cook, Cortese, D'Silva, Grootes, Holwerda, Hopkins, Jarvis, Lidman,
  Phillipps, \& Siudek}]{davies_deep_2022}
Davies, L. J.~M., Thorne, J.~E., Bellstedt, S., {et~al.} 2022, Monthly Notices
  of the Royal Astronomical Society, 509, 4392, \dodoi{10.1093/mnras/stab3145}

\bibitem[{Driver {et~al.}(2018)Driver, Andrews, Da~Cunha, Davies, Lagos,
  Robotham, Vinsen, Wright, Alpaslan, Bland-Hawthorn, Bourne, Brough, Bremer,
  Cluver, Colless, Conselice, Dunne, Eales, Gomez, Holwerda, Hopkins, Kafle,
  Kelvin, Loveday, Liske, Maddox, Phillipps, Pimbblet, Rowlands, Sansom,
  Taylor, Wang, \& Wilkins}]{driver_gamag10-cosmos3d-hst_2018}
Driver, S., Andrews, S.~K., Da~Cunha, E., {et~al.} 2018, Monthly Notices of the
  Royal Astronomical Society: Letters, 475, 2891, \dodoi{10.1093/mnras/stx2728}

\bibitem[{D'Silva {et~al.}(2023{\natexlab{a}})D'Silva, Lagos, Davies, Lovell,
  \& Vijayan}]{dsilva_unveiling_2023}
D'Silva, J. C.~J., Lagos, C. D.~P., Davies, L. J.~M., Lovell, C.~C., \&
  Vijayan, A.~P. 2023{\natexlab{a}}, Monthly Notices of the Royal Astronomical
  Society, 518, 456, \dodoi{10.1093/mnras/stac2878}

\bibitem[{D'Silva {et~al.}(2023{\natexlab{b}})D'Silva, Driver, Lagos, Robotham,
  Bellstedt, Davies, Thorne, Bland-Hawthorn, Bravo, Holwerda, Phillipps,
  Seymour, Siudek, \& Windhorst}]{dsilva_gamadevils_2023}
D'Silva, J. C.~J., Driver, S.~P., Lagos, C. D.~P., {et~al.} 2023{\natexlab{b}},
  Monthly Notices of the Royal Astronomical Society, 524, 1448,
  \dodoi{10.1093/mnras/stad1974}

\bibitem[{{D'Silva et al.}(2023)}]{jumprope}
{D'Silva et al.} 2023, {JordanDSilva}/{JUMPROPE}: {Version} 1.0.0,  Zenodo,
  \dodoi{10.5281/zenodo.10148539}

\bibitem[{Ferrarese \& Merritt(2000)}]{ferrarese_fundamental_2000}
Ferrarese, L., \& Merritt, D. 2000, The Astrophysical Journal, 539, L9,
  \dodoi{10.1086/312838}

\bibitem[{{Ferreira} {et~al.}(2023){Ferreira}, {Conselice}, {Sazonova},
  {Ferrari}, {Caruana}, {Tohill}, {Lucatelli}, {Adams}, {Irodotou}, {Marshall},
  {Roper}, {Lovell}, {Verma}, {Austin}, {Trussler}, \&
  {Wilkins}}]{ferreira_jwst_2022}
{Ferreira}, L., {Conselice}, C.~J., {Sazonova}, E., {et~al.} 2023, \apj, 955,
  94, \dodoi{10.3847/1538-4357/acec76}

\bibitem[{Finkelstein \& Bagley(2022)}]{finkelstein_coevolution_2022}
Finkelstein, S.~L., \& Bagley, M.~B. 2022, The Astrophysical Journal, 938, 25,
  \dodoi{10.3847/1538-4357/ac89eb}

\bibitem[{Finkelstein {et~al.}(2023)Finkelstein, Bagley, Ferguson, Wilkins,
  Kartaltepe, Papovich, Yung, Haro, Behroozi, Dickinson, Kocevski, Koekemoer,
  Larson, Le~Bail, Morales, Pérez-González, Burgarella, Davé, Hirschmann,
  Somerville, Wuyts, Bromm, Casey, Fontana, Fujimoto, Gardner, Giavalisco,
  Grazian, Grogin, Hathi, Hutchison, Jha, Jogee, Kewley, Kirkpatrick, Long,
  Lotz, Pentericci, Pierel, Pirzkal, Ravindranath, Ryan, Trump, Yang,
  Bhatawdekar, Bisigello, Buat, Calabrò, Castellano, Cleri, Cooper, Croton,
  Daddi, Dekel, Elbaz, Franco, Gawiser, Holwerda, Huertas-Company, Jaskot,
  Leung, Lucas, Mobasher, Pandya, Tacchella, Weiner, \&
  Zavala}]{finkelstein_ceers_2023}
Finkelstein, S.~L., Bagley, M.~B., Ferguson, H.~C., {et~al.} 2023, The
  Astrophysical Journal, 946, L13, \dodoi{10.3847/2041-8213/acade4}

\bibitem[{Fritz {et~al.}(2006)Fritz, Franceschini, \&
  Hatziminaoglou}]{fritz06agnmodel}
Fritz, J., Franceschini, A., \& Hatziminaoglou, E. 2006, {\textbackslash}mnras,
  366, 767, \dodoi{10.1111/j.1365-2966.2006.09866.x}

\bibitem[{Gardner {et~al.}(2006)Gardner, Mather, Clampin, Doyon, Greenhouse,
  Hammel, Hutchings, Jakobsen, Lilly, Long, Lunine, McCaughrean, Mountain,
  Nella, Rieke, Rieke, Rix, Smith, Sonneborn, Stiavelli, Stockman, Windhorst,
  \& Wright}]{gardner_james_2006}
Gardner, J.~P., Mather, J.~C., Clampin, M., {et~al.} 2006, Space Science
  Reviews, 123, 485, \dodoi{10.1007/s11214-006-8315-7}

\bibitem[{Glazebrook {et~al.}(2023)Glazebrook, Nanayakkara, Schreiber, Lagos,
  Kawinwanichakij, Jacobs, Chittenden, Brammer, Kacprzak, Labbe, Marchesini,
  Marsan, Oesch, Papovich, Remus, Tran, Esdaile, \&
  Chandro~Gomez}]{glazebrook_extraordinarily_2023}
Glazebrook, K., Nanayakkara, T., Schreiber, C., {et~al.} 2023, An
  extraordinarily massive galaxy that formed its stars at \$z rsim 11\$,
  \dodoi{10.48550/arXiv.2308.05606}

\bibitem[{Harikane {et~al.}(2023{\natexlab{a}})Harikane, Nakajima, Ouchi,
  Umeda, Isobe, Ono, Xu, \& Zhang}]{harikane_pure_2023}
Harikane, Y., Nakajima, K., Ouchi, M., {et~al.} 2023{\natexlab{a}}, Pure
  {Spectroscopic} {Constraints} on {UV} {Luminosity} {Functions} and {Cosmic}
  {Star} {Formation} {History} {From} 25 {Galaxies} at
  \$z\_{\textbackslash}mathrm\{spec\}=8.61-13.20\$ {Confirmed} with
  {JWST}/{NIRSpec}, Tech. rep., \dodoi{10.48550/arXiv.2304.06658}

\bibitem[{Harikane {et~al.}(2022)Harikane, Ono, Ouchi, Liu, Sawicki, Shibuya,
  Behroozi, He, Shimasaku, Arnouts, Coupon, Fujimoto, Gwyn, Huang, Inoue,
  Kashikawa, Komiyama, Matsuoka, \& Willott}]{harikane_goldrush_2022}
Harikane, Y., Ono, Y., Ouchi, M., {et~al.} 2022, The Astrophysical Journal
  Supplement Series, 259, 20, \dodoi{10.3847/1538-4365/ac3dfc}

\bibitem[{Harikane {et~al.}(2023{\natexlab{b}})Harikane, Ouchi, Oguri, Ono,
  Nakajima, Isobe, Umeda, Mawatari, \& Zhang}]{harikane_comprehensive_2023}
Harikane, Y., Ouchi, M., Oguri, M., {et~al.} 2023{\natexlab{b}}, The
  Astrophysical Journal Supplement Series, 265, 5,
  \dodoi{10.3847/1538-4365/acaaa9}

\bibitem[{Harikane {et~al.}(2023{\natexlab{c}})Harikane, Zhang, Nakajima,
  Ouchi, Isobe, Ono, Hatano, Xu, \& Umeda}]{harikane_jwstnirspec_2023}
Harikane, Y., Zhang, Y., Nakajima, K., {et~al.} 2023{\natexlab{c}},
  {JWST}/{NIRSpec} {First} {Census} of {Broad}-{Line} {AGNs} at z=4-7:
  {Detection} of 10 {Faint} {AGNs} with
  {M}\_BH{\textasciitilde}10{\textasciicircum}6-10{\textasciicircum}7 {M}\_sun
  and {Their} {Host} {Galaxy} {Properties}, Tech. rep.,
  \dodoi{10.48550/arXiv.2303.11946}

\bibitem[{{Heintz} {et~al.}(2023){Heintz}, {Brammer}, {Gim{\'e}nez-Arteaga},
  {Strait}, {del P. Lagos}, {Vijayan}, {Matthee}, {Watson}, {Mason}, {Hutter},
  {Toft}, {Fynbo}, \& {Oesch}}]{heintz_fundamental_2022}
{Heintz}, K.~E., {Brammer}, G.~B., {Gim{\'e}nez-Arteaga}, C., {et~al.} 2023,
  Nature Astronomy, \dodoi{10.1038/s41550-023-02078-7}

\bibitem[{Juodžbalis {et~al.}(2023)Juodžbalis, Conselice, Singh, Adams,
  Ormerod, Harvey, Austin, Volonteri, Cohen, Jansen, Summers, Windhorst,
  D'Silva, Koekemoer, Coe, Driver, Frye, Grogin, Marshall, Nonino, Pirzkal,
  Robotham, Ryan, Ortiz~III, Tompkins, Willmer, \&
  Yan}]{juodzbalis_epochs_2023}
Juodžbalis, I., Conselice, C.~J., Singh, M., {et~al.} 2023, Monthly Notices of
  the Royal Astronomical Society, 525, 1353, \dodoi{10.1093/mnras/stad2396}

\bibitem[{Katsianis {et~al.}(2019)Katsianis, Zheng, Gonzalez, Blanc, Lagos,
  Davies, Camps, Trčka, Baes, Schaye, Trayford, Theuns, \&
  Stalevski}]{katsianis_evolving_2019}
Katsianis, A., Zheng, X., Gonzalez, V., {et~al.} 2019, The Astrophysical
  Journal, 879, 11, \dodoi{10.3847/1538-4357/ab1f8d}

\bibitem[{Kauffmann \& Haehnelt(2000)}]{kauffmann_unified_2000}
Kauffmann, G., \& Haehnelt, M. 2000, Monthly Notices of the Royal Astronomical
  Society, 311, 576, \dodoi{10.1046/j.1365-8711.2000.03077.x}

\bibitem[{Kirby {et~al.}(2013)Kirby, Boylan-Kolchin, Cohen, Geha, Bullock, \&
  Kaplinghat}]{kirby_segue_2013}
Kirby, E.~N., Boylan-Kolchin, M., Cohen, J.~G., {et~al.} 2013, The
  Astrophysical Journal, 770, 16, \dodoi{10.1088/0004-637X/770/1/16}

\bibitem[{{Kocevski} {et~al.}(2023){Kocevski}, {Onoue}, {Inayoshi}, {Trump},
  {Arrabal Haro}, {Grazian}, {Dickinson}, {Finkelstein}, {Kartaltepe},
  {Hirschmann}, {Aird}, {Holwerda}, {Fujimoto}, {Juneau}, {Amor{\'\i}n},
  {Backhaus}, {Bagley}, {Barro}, {Bell}, {Bisigello}, {Calabr{\`o}}, {Cleri},
  {Cooper}, {Ding}, {Grogin}, {Ho}, {Hutchison}, {Inoue}, {Jiang}, {Jones},
  {Koekemoer}, {Li}, {Li}, {McGrath}, {Molina}, {Papovich},
  {P{\'e}rez-Gonz{\'a}lez}, {Pirzkal}, {Wilkins}, {Yang}, \&
  {Yung}}]{kocevski_hidden_2023}
{Kocevski}, D.~D., {Onoue}, M., {Inayoshi}, K., {et~al.} 2023, \apjl, 954, L4,
  \dodoi{10.3847/2041-8213/ace5a0}

\bibitem[{Koekemoer {et~al.}(2011)Koekemoer, Faber, Ferguson, Grogin, Kocevski,
  Koo, Lai, Lotz, Lucas, McGrath, Ogaz, Rajan, Riess, Rodney, Strolger,
  Casertano, Castellano, Dahlen, Dickinson, Dolch, Fontana, Giavalisco,
  Grazian, Guo, Hathi, Huang, van~der Wel, Yan, Acquaviva, Alexander, Almaini,
  Ashby, Barden, Bell, Bournaud, Brown, Caputi, Cassata, Challis, Chary,
  Cheung, Cirasuolo, Conselice, Roshan~Cooray, Croton, Daddi, Davé, de~Mello,
  de~Ravel, Dekel, Donley, Dunlop, Dutton, Elbaz, Fazio, Filippenko,
  Finkelstein, Frazer, Gardner, Garnavich, Gawiser, Gruetzbauch, Hartley,
  Häussler, Herrington, Hopkins, Huang, Jha, Johnson, Kartaltepe, Khostovan,
  Kirshner, Lani, Lee, Li, Madau, McCarthy, McIntosh, McLure, McPartland,
  Mobasher, Moreira, Mortlock, Moustakas, Mozena, Nandra, Newman, Nielsen,
  Niemi, Noeske, Papovich, Pentericci, Pope, Primack, Ravindranath, Reddy,
  Renzini, Rix, Robaina, Rosario, Rosati, Salimbeni, Scarlata, Siana, Simard,
  Smidt, Snyder, Somerville, Spinrad, Straughn, Telford, Teplitz, Trump,
  Vargas, Villforth, Wagner, Wandro, Wechsler, Weiner, Wiklind, Wild, Wilson,
  Wuyts, \& Yun}]{koekemoer_candels_2011}
Koekemoer, A.~M., Faber, S.~M., Ferguson, H.~C., {et~al.} 2011, The
  Astrophysical Journal Supplement Series, 197, 36,
  \dodoi{10.1088/0067-0049/197/2/36}

\bibitem[{{Kroupa} {et~al.}(2020){Kroupa}, {Subr}, {Jerabkova}, \&
  {Wang}}]{kroupa_very_2020}
{Kroupa}, P., {Subr}, L., {Jerabkova}, T., \& {Wang}, L. 2020, \mnras, 498,
  5652, \dodoi{10.1093/mnras/staa2276}

\bibitem[{Lagos {et~al.}(2018)Lagos, Tobar, Robotham, Obreschkow, Mitchell,
  Power, \& Elahi}]{lagos_shark_2018}
Lagos, C. d.~P., Tobar, R.~J., Robotham, A. S.~G., {et~al.} 2018, Monthly
  Notices of the Royal Astronomical Society, 481, 3573,
  \dodoi{10.1093/mnras/sty2440}

\bibitem[{Lagos {et~al.}(2023)Lagos, Bravo, Tobar, Obreschkow, Power, Robotham,
  Proctor, Hansen, Chandro-Gomez, \& Carrivick}]{lagos_quenching_2023}
Lagos, C. D.~P., Bravo, M., Tobar, R., {et~al.} 2023, Quenching massive
  galaxies across cosmic time with the semi-analytic model {SHARK} v2.0,
  \dodoi{10.48550/arXiv.2309.02310}

\bibitem[{{Larson} {et~al.}(2023){Larson}, {Finkelstein}, {Kocevski},
  {Hutchison}, {Trump}, {Arrabal Haro}, {Bromm}, {Cleri}, {Dickinson},
  {Fujimoto}, {Kartaltepe}, {Koekemoer}, {Papovich}, {Pirzkal}, {Tacchella},
  {Zavala}, {Bagley}, {Behroozi}, {Champagne}, {Cole}, {Jung}, {Morales},
  {Yang}, {Zhang}, {Zitrin}, {Amor{\'\i}n}, {Burgarella}, {Casey}, {Ch{\'a}vez
  Ortiz}, {Cox}, {Chworowsky}, {Fontana}, {Gawiser}, {Grazian}, {Grogin},
  {Harish}, {Hathi}, {Hirschmann}, {Holwerda}, {Juneau}, {Leung}, {Lucas},
  {McGrath}, {P{\'e}rez-Gonz{\'a}lez}, {Rigby}, {Seill{\'e}}, {Simons}, {de La
  Vega}, {Weiner}, {Wilkins}, {Yung}, \& {Ceers Team}}]{larson_ceers_2023}
{Larson}, R.~L., {Finkelstein}, S.~L., {Kocevski}, D.~D., {et~al.} 2023, \apjl,
  953, L29, \dodoi{10.3847/2041-8213/ace619}

\bibitem[{Levesque {et~al.}(2010)Levesque, Kewley, \&
  Larson}]{levesque_theoretical_2010}
Levesque, E.~M., Kewley, L.~J., \& Larson, K.~L. 2010, The Astronomical
  Journal, 139, 712, \dodoi{10.1088/0004-6256/139/2/712}

\bibitem[{Lilly {et~al.}(1996)Lilly, Le~Fevre, Hammer, \&
  Crampton}]{lilly_canada-france_1996}
Lilly, S.~J., Le~Fevre, O., Hammer, F., \& Crampton, D. 1996, The Astrophysical
  Journal, 460, L1, \dodoi{10.1086/309975}

\bibitem[{Lotz {et~al.}(2017)Lotz, Koekemoer, Coe, Grogin, Capak, Mack,
  Anderson, Avila, Barker, Borncamp, Brammer, Durbin, Gunning, Hilbert,
  Jenkner, Khandrika, Levay, Lucas, MacKenty, Ogaz, Porterfield, Reid,
  Robberto, Royle, Smith, Storrie-Lombardi, Sunnquist, Surace, Taylor,
  Williams, Bullock, Dickinson, Finkelstein, Natarajan, Richard, Robertson,
  Tumlinson, Zitrin, Flanagan, Sembach, Soifer, \&
  Mountain}]{lotz_frontier_2017}
Lotz, J.~M., Koekemoer, A., Coe, D., {et~al.} 2017, The Astrophysical Journal,
  837, 97, \dodoi{10.3847/1538-4357/837/1/97}

\bibitem[{{Lovell} {et~al.}(2023{\natexlab{a}}){Lovell}, {Harrison},
  {Harikane}, {Tacchella}, \& {Wilkins}}]{lovell_extreme_2022}
{Lovell}, C.~C., {Harrison}, I., {Harikane}, Y., {Tacchella}, S., \& {Wilkins},
  S.~M. 2023{\natexlab{a}}, \mnras, 518, 2511, \dodoi{10.1093/mnras/stac3224}

\bibitem[{Lovell {et~al.}(2021)Lovell, Vijayan, Thomas, Wilkins, Barnes,
  Irodotou, \& Roper}]{lovell_first_2021}
Lovell, C.~C., Vijayan, A.~P., Thomas, P.~A., {et~al.} 2021, Monthly Notices of
  the Royal Astronomical Society, 500, 2127, \dodoi{10.1093/mnras/staa3360}

\bibitem[{{Lovell} {et~al.}(2023{\natexlab{b}}){Lovell}, {Roper}, {Vijayan},
  {Seeyave}, {Irodotou}, {Wilkins}, {Conselice}, {Fortuni}, {Kuusisto},
  {Merlin}, {Santini}, \& {Thomas}}]{lovell_first_2023}
{Lovell}, C.~C., {Roper}, W., {Vijayan}, A.~P., {et~al.} 2023{\natexlab{b}},
  \mnras, 525, 5520, \dodoi{10.1093/mnras/stad2550}

\bibitem[{Madau \& Dickinson(2014)}]{madau_cosmic_2014}
Madau, P., \& Dickinson, M. 2014, Annual Review of Astronomy and Astrophysics,
  vol. 52, p.415-486, 52, 415, \dodoi{10.1146/annurev-astro-081811-125615}

\bibitem[{Magorrian {et~al.}(1998)Magorrian, Tremaine, Richstone, Bender,
  Bower, Dressler, Faber, Gebhardt, Green, Grillmair, Kormendy, \&
  Lauer}]{magorrian_demography_1998}
Magorrian, J., Tremaine, S., Richstone, D., {et~al.} 1998, The Astronomical
  Journal, 115, 2285, \dodoi{10.1086/300353}

\bibitem[{Maiolino {et~al.}(2023)Maiolino, Scholtz, Curtis-Lake, Carniani,
  Baker, de~Graaff, Tacchella, Übler, D'Eugenio, Witstok, Curti, Arribas,
  Bunker, Charlot, Chevallard, Eisenstein, Egami, Ji, Jones, Lyu, Rawle,
  Robertson, Rujopakarn, Perna, Sun, Venturi, Williams, \&
  Willott}]{maiolino_jades_2023}
Maiolino, R., Scholtz, J., Curtis-Lake, E., {et~al.} 2023, {JADES}. {The}
  diverse population of infant {Black} {Holes} at 4,
  \dodoi{10.48550/arXiv.2308.01230}

\bibitem[{Matthee \& Schaye(2019)}]{matthee_origin_2019}
Matthee, J., \& Schaye, J. 2019, Monthly Notices of the Royal Astronomical
  Society, 484, 915, \dodoi{10.1093/mnras/stz030}

\bibitem[{Naidu {et~al.}(2022)Naidu, Oesch, van Dokkum, Nelson, Suess, Brammer,
  Whitaker, Illingworth, Bouwens, Tacchella, Matthee, Allen, Bezanson, Conroy,
  Labbe, Leja, Leonova, Magee, Price, Setton, Strait, Stefanon, Toft, Weaver,
  \& Weibel}]{naidu_two_2022}
Naidu, R.~P., Oesch, P.~A., van Dokkum, P., {et~al.} 2022, The Astrophysical
  Journal, 940, L14, \dodoi{10.3847/2041-8213/ac9b22}

\bibitem[{{Nakajima} {et~al.}(2023){Nakajima}, {Ouchi}, {Isobe}, {Harikane},
  {Zhang}, {Ono}, {Umeda}, \& {Oguri}}]{nakajima_jwst_2023}
{Nakajima}, K., {Ouchi}, M., {Isobe}, Y., {et~al.} 2023, \apjs, 269, 33,
  \dodoi{10.3847/1538-4365/acd556}

\bibitem[{Noeske {et~al.}(2007)Noeske, Weiner, Faber, Papovich, Koo,
  Somerville, Bundy, Conselice, Newman, Schiminovich, Le~Floc'h, Coil, Rieke,
  Lotz, Primack, Barmby, Cooper, Davis, Ellis, Fazio, Guhathakurta, Huang,
  Kassin, Martin, Phillips, Rich, Small, Willmer, \& Wilson}]{noeske_star_2007}
Noeske, K.~G., Weiner, B.~J., Faber, S.~M., {et~al.} 2007, The Astrophysical
  Journal, 660, L43, \dodoi{10.1086/517926}

\bibitem[{{Ormerod} {et~al.}(2023){Ormerod}, {Conselice}, {Adams}, {Harvey},
  {Austin}, {Trussler}, {Ferreira}, {Caruana}, {Lucatelli}, {Li}, \&
  {Roper}}]{ormerod_epochs_2023}
{Ormerod}, K., {Conselice}, C.~J., {Adams}, N.~J., {et~al.} 2023, \mnras,
  \dodoi{10.1093/mnras/stad3597}

\bibitem[{Richards {et~al.}(2006)Richards, Strauss, Fan, Hall, Jester,
  Schneider, Vanden~Berk, Stoughton, Anderson, Brunner, Gray, Gunn, Ivezić,
  Kirkland, Knapp, Loveday, Meiksin, Pope, Szalay, Thakar, Yanny, York,
  Barentine, Brewington, Brinkmann, Fukugita, Harvanek, Kent, Kleinman,
  Krzesiński, Long, Lupton, Nash, Neilsen, Nitta, Schlegel, \&
  Snedden}]{richards_sloan_2006}
Richards, G.~T., Strauss, M.~A., Fan, X., {et~al.} 2006, The Astronomical
  Journal, 131, 2766, \dodoi{10.1086/503559}

\bibitem[{{Rieke} {et~al.}(2005){Rieke}, {Kelly}, \&
  {Horner}}]{rieke_overview_2005}
{Rieke}, M.~J., {Kelly}, D., \& {Horner}, S. 2005, in Society of Photo-Optical
  Instrumentation Engineers (SPIE) Conference Series, Vol. 5904, Cryogenic
  Optical Systems and Instruments XI, ed. J.~B. {Heaney} \& L.~G. {Burriesci},
  1--8, \dodoi{10.1117/12.615554}

\bibitem[{Rieke {et~al.}(2023)Rieke, Kelly, Misselt, Stansberry, Boyer, Beatty,
  Egami, Florian, Greene, Hainline, Leisenring, Roellig, Schlawin, Sun, Tinnin,
  Williams, Willmer, Wilson, Clark, Rohrbach, Brooks, Canipe, Correnti,
  DiFelice, Gennaro, Girard, Hartig, Hilbert, Koekemoer, Nikolov, Pirzkal,
  Rest, Robberto, Sunnquist, Telfer, Wu, Ferry, Lewis, Baum, Beichman, Doyon,
  Dressler, Eisenstein, Ferrarese, Hodapp, Horner, Jaffe, Johnstone, Krist,
  Martin, McCarthy, Meyer, Rieke, Trauger, \& Young}]{rieke_performance_2023}
Rieke, M.~J., Kelly, D.~M., Misselt, K., {et~al.} 2023, Publications of the
  Astronomical Society of the Pacific, 135, 028001,
  \dodoi{10.1088/1538-3873/acac53}

\bibitem[{Robotham {et~al.}(2020)Robotham, Bellstedt, Lagos, Thorne, Davies,
  Driver, \& Bravo}]{robotham_prospect_2020}
Robotham, A. S.~G., Bellstedt, S., Lagos, C. d.~P., {et~al.} 2020, Monthly
  Notices of the Royal Astronomical Society, 495, 905,
  \dodoi{10.1093/mnras/staa1116}

\bibitem[{Robotham {et~al.}(2018)Robotham, Davies, Driver, Koushan, Taranu,
  Casura, \& Liske}]{robotham_profound_2018}
Robotham, A. S.~G., Davies, L. J.~M., Driver, S.~P., {et~al.} 2018, Monthly
  Notices of the Royal Astronomical Society, 476, 3137,
  \dodoi{10.1093/mnras/sty440}

\bibitem[{Robotham {et~al.}(2023{\natexlab{a}})Robotham, D'Silva, Windhorst,
  Jansen, Summers, Driver, Wilmer, \& Bellstedt}]{robotham_dynamic_2023}
Robotham, A. S.~G., D'Silva, J. C.~J., Windhorst, R.~A., {et~al.}
  2023{\natexlab{a}}, Publications of the Astronomical Society of the Pacific,
  135, 085003, \dodoi{10.1088/1538-3873/acea42}

\bibitem[{Robotham \& Obreschkow(2015)}]{robotham_hyper-fit_2015}
Robotham, A. S.~G., \& Obreschkow, D. 2015, Publications of the Astronomical
  Society of Australia, 32, e033, \dodoi{10.1017/pasa.2015.33}

\bibitem[{Robotham {et~al.}(2023{\natexlab{b}})Robotham, Tobar, Bellstedt,
  Casura, Cook, D'Silva, Davies, Driver, Li, \&
  Garate-Nuñez}]{robotham_propane_2023}
Robotham, A. S.~G., Tobar, R., Bellstedt, S., {et~al.} 2023{\natexlab{b}},
  {ProPane}: {Image} {Warping} with {Fire},  arXiv.
\newblock \url{http://arxiv.org/abs/2311.01761}

\bibitem[{Salim {et~al.}(2007)Salim, Rich, Charlot, Brinchmann, Johnson,
  Schiminovich, Seibert, Mallery, Heckman, Forster, Friedman, Martin,
  Morrissey, Neff, Small, Wyder, Bianchi, Donas, Lee, Madore, Milliard, Szalay,
  Welsh, \& Yi}]{salim_uv_2007}
Salim, S., Rich, R.~M., Charlot, S., {et~al.} 2007, The Astrophysical Journal
  Supplement Series, 173, 267, \dodoi{10.1086/519218}

\bibitem[{Schaye {et~al.}(2010)Schaye, Dalla~Vecchia, Booth, Wiersma, Theuns,
  Haas, Bertone, Duffy, McCarthy, \& van~de Voort}]{schaye_physics_2010}
Schaye, J., Dalla~Vecchia, C., Booth, C.~M., {et~al.} 2010, Monthly Notices of
  the Royal Astronomical Society, 402, 1536,
  \dodoi{10.1111/j.1365-2966.2009.16029.x10.48550/arXiv.0909.5196}

\bibitem[{Schechter(1976)}]{schechter_analytic_1976}
Schechter, P. 1976, The Astrophysical Journal, 203, 297, \dodoi{10.1086/154079}

\bibitem[{{Schneider} {et~al.}(2023){Schneider}, {Valiante}, {Trinca},
  {Graziani}, {Volonteri}, \& {Maiolino}}]{schneider_are_2023}
{Schneider}, R., {Valiante}, R., {Trinca}, A., {et~al.} 2023, \mnras, 526,
  3250, \dodoi{10.1093/mnras/stad2503}

\bibitem[{Shen {et~al.}(2020)Shen, Hopkins, Faucher-Giguère, Alexander,
  Richards, Ross, \& Hickox}]{shen_bolometric_2020}
Shen, X., Hopkins, P.~F., Faucher-Giguère, C.-A., {et~al.} 2020, Monthly
  Notices of the Royal Astronomical Society, 495, 3252,
  \dodoi{10.1093/mnras/staa1381}

\bibitem[{Song {et~al.}(2016)Song, Finkelstein, Ashby, Grazian, Lu, Papovich,
  Salmon, Somerville, Dickinson, Duncan, Faber, Fazio, Ferguson, Fontana, Guo,
  Hathi, Lee, Merlin, \& Willner}]{song_evolution_2016}
Song, M., Finkelstein, S.~L., Ashby, M. L.~N., {et~al.} 2016, The Astrophysical
  Journal, 825, 5, \dodoi{10.3847/0004-637X/825/1/5}

\bibitem[{Speagle {et~al.}(2014)Speagle, Steinhardt, Capak, \&
  Silverman}]{speagle_highly_2014}
Speagle, J.~S., Steinhardt, C.~L., Capak, P.~L., \& Silverman, J.~D. 2014, The
  Astrophysical Journal Supplement Series, 214, 15,
  \dodoi{10.1088/0067-0049/214/2/15}

\bibitem[{Stefanon {et~al.}(2021)Stefanon, Bouwens, Labbé, Illingworth,
  Gonzalez, \& Oesch}]{stefanon_galaxy_2021}
Stefanon, M., Bouwens, R.~J., Labbé, I., {et~al.} 2021, ApJ, 922, 29,
  \dodoi{10.3847/1538-4357/ac1bb6}

\bibitem[{Tacchella {et~al.}(2016)Tacchella, Dekel, Carollo, Ceverino, DeGraf,
  Lapiner, Mandelker, \& Primack~Joel}]{tacchella_confinement_2016}
Tacchella, S., Dekel, A., Carollo, C.~M., {et~al.} 2016, Monthly Notices of the
  Royal Astronomical Society, 457, 2790, \dodoi{10.1093/mnras/stw131}

\bibitem[{Tacchella {et~al.}(2023)Tacchella, Eisenstein, Hainline, Johnson,
  Baker, Helton, Robertson, Suess, Chen, Nelson, Puskás, Sun, Alberts, Egami,
  Hausen, Rieke, Rieke, Shivaei, Williams, Willmer, Bunker, Cameron, Carniani,
  Charlot, Curti, Curtis-Lake, Looser, Maiolino, Maseda, Rawle, Rix, Smit,
  Übler, Willott, Witstok, Baum, Bhatawdekar, Boyett, Danhaive, de~Graaff,
  Endsley, Ji, Lyu, Sandles, Saxena, Scholtz, Topping, \&
  Whitler}]{tacchella_jades_2023}
Tacchella, S., Eisenstein, D.~J., Hainline, K., {et~al.} 2023, The
  Astrophysical Journal, 952, 74, \dodoi{10.3847/1538-4357/acdbc6}

\bibitem[{Thorne {et~al.}(2021)Thorne, Robotham, Davies, Bellstedt, Driver,
  Bravo, Bremer, Holwerda, Hopkins, Lagos, Phillipps, Siudek, Taylor, \&
  Wright}]{thorne_deep_2021}
Thorne, J.~E., Robotham, A. S.~G., Davies, L. J.~M., {et~al.} 2021, Monthly
  Notices of the Royal Astronomical Society, 505, 540,
  \dodoi{10.1093/mnras/stab1294}

\bibitem[{Thorne {et~al.}(2022)Thorne, Robotham, Davies, Bellstedt, Brown,
  Croom, Delvecchio, Groves, Jarvis, Shabala, Seymour, Whittam, Bravo, Cook,
  Driver, Holwerda, Phillipps, \& Siudek}]{thorne_deep_2022}
---. 2022, Monthly Notices of the Royal Astronomical Society, 509, 4940,
  \dodoi{10.1093/mnras/stab3208}

\bibitem[{Treu {et~al.}(2022)Treu, Roberts-Borsani, Bradac, Brammer, Fontana,
  Henry, Mason, Morishita, Pentericci, Wang, Acebron, Bagley, Bergamini,
  Belfiori, Bonchi, Boyett, Boutsia, Calabró, Caminha, Castellano, Dressler,
  Glazebrook, Grillo, Jacobs, Jones, Kelly, Leethochawalit, Malkan, Marchesini,
  Mascia, Mercurio, Merlin, Nanayakkara, Nonino, Paris, Poggianti, Rosati,
  Santini, Scarlata, Shipley, Strait, Trenti, Tubthong, Vanzella, Vulcani, \&
  Yang}]{treu_glass-jwst_2022}
Treu, T., Roberts-Borsani, G., Bradac, M., {et~al.} 2022, The Astrophysical
  Journal, 935, 110, \dodoi{10.3847/1538-4357/ac8158}

\bibitem[{{{\"U}bler} {et~al.}(2023){{\"U}bler}, {Maiolino}, {Curtis-Lake},
  {P{\'e}rez-Gonz{\'a}lez}, {Curti}, {Perna}, {Arribas}, {Charlot}, {Marshall},
  {D'Eugenio}, {Scholtz}, {Bunker}, {Carniani}, {Ferruit}, {Jakobsen}, {Rix},
  {Rodr{\'\i}guez Del Pino}, {Willott}, {Boeker}, {Cresci}, {Jones}, {Kumari},
  \& {Rawle}}]{ubler_ga-nifs_2023}
{{\"U}bler}, H., {Maiolino}, R., {Curtis-Lake}, E., {et~al.} 2023, \aap, 677,
  A145, \dodoi{10.1051/0004-6361/202346137}

\bibitem[{Vijayan {et~al.}(2021)Vijayan, Lovell, Wilkins, Thomas, Barnes,
  Irodotou, Kuusisto, \& Roper}]{vijayan_first_2021}
Vijayan, A.~P., Lovell, C.~C., Wilkins, S.~M., {et~al.} 2021, Monthly Notices
  of the Royal Astronomical Society, 501, 3289, \dodoi{10.1093/mnras/staa3715}

\bibitem[{{Weaver} {et~al.}(2023){Weaver}, {Davidzon}, {Toft}, {Ilbert},
  {McCracken}, {Gould}, {Jespersen}, {Steinhardt}, {Lagos}, {Capak}, {Casey},
  {Chartab}, {Faisst}, {Hayward}, {Kartaltepe}, {Kauffmann}, {Koekemoer},
  {Kokorev}, {Laigle}, {Liu}, {Long}, {Magdis}, {McPartland}, {Milvang-Jensen},
  {Mobasher}, {Moneti}, {Peng}, {Sanders}, {Shuntov}, {Sneppen}, {Valentino},
  {Zalesky}, \& {Zamorani}}]{weaver_cosmos2020_2023}
{Weaver}, J.~R., {Davidzon}, I., {Toft}, S., {et~al.} 2023, \aap, 677, A184,
  \dodoi{10.1051/0004-6361/202245581}

\bibitem[{Weigel {et~al.}(2016)Weigel, Schawinski, \&
  Bruderer}]{weigel_stellar_2016}
Weigel, A.~K., Schawinski, K., \& Bruderer, C. 2016, Monthly Notices of the
  Royal Astronomical Society, 459, 2150, \dodoi{10.1093/mnras/stw756}

\bibitem[{{Windhorst} {et~al.}(2023){Windhorst}, {Cohen}, {Jansen}, {Summers},
  {Tompkins}, {Conselice}, {Driver}, {Yan}, {Coe}, {Frye}, {Grogin},
  {Koekemoer}, {Marshall}, {O'Brien}, {Pirzkal}, {Robotham}, {Ryan}, {Willmer},
  {Carleton}, {Diego}, {Keel}, {Porto}, {Redshaw}, {Scheller}, {Wilkins},
  {Willner}, {Zitrin}, {Adams}, {Austin}, {Arendt}, {Beacom}, {Bhatawdekar},
  {Bradley}, {Broadhurst}, {Cheng}, {Civano}, {Dai}, {Dole}, {D'Silva},
  {Duncan}, {Fazio}, {Ferrami}, {Ferreira}, {Finkelstein}, {Furtak}, {Gim},
  {Griffiths}, {Hammel}, {Harrington}, {Hathi}, {Holwerda}, {Honor}, {Huang},
  {Hyun}, {Im}, {Joshi}, {Kamieneski}, {Kelly}, {Larson}, {Li}, {Lim}, {Ma},
  {Maksym}, {Manzoni}, {Meena}, {Milam}, {Nonino}, {Pascale}, {Petric},
  {Pierel}, {del Carmen Polletta}, {R{\"o}ttgering}, {Rutkowski}, {Smail},
  {Straughn}, {Strolger}, {Swirbul}, {Trussler}, {Wang}, {Welch}, {B. Wyithe},
  {Yun}, {Zackrisson}, {Zhang}, \& {Zhao}}]{windhorst_webbs_2022}
{Windhorst}, R.~A., {Cohen}, S.~H., {Jansen}, R.~A., {et~al.} 2023, \aj, 165,
  13, \dodoi{10.3847/1538-3881/aca163}

\bibitem[{Yan {et~al.}(2023)Yan, Cohen, Windhorst, Jansen, Ma, Beacom, Ling,
  Cheng, Huang, Grogin, Willner, Yun, Hammel, Milam, Conselice, Driver, Frye,
  Marshall, Koekemoer, Willmer, Robotham, D'Silva, Summers, Lim, Harrington,
  Ferreira, Diego, Pirzkal, Wilkins, Wang, Hathi, Zitrin, Bhatawdekar, Adams,
  Furtak, Maksym, Rutkowski, \& Fazio}]{yan_jwsts_2023}
Yan, H., Cohen, S.~H., Windhorst, R.~A., {et~al.} 2023, The Astrophysical
  Journal, 942, L8, \dodoi{10.3847/2041-8213/aca974}

\end{thebibliography}
\bibliographystyle{aasjournal}

\end{document}